\documentclass[10pt,twocolumn,journal]{IEEEtran}
\usepackage{dsfont}
\usepackage{setspace}
\usepackage{clrscode}
\usepackage{pict2e}
\usepackage[font=small,labelfont=small]{caption}

\usepackage[T1]{fontenc}
\usepackage[utf8]{inputenc}
\usepackage{authblk}
\usepackage{blindtext}
\usepackage{makecell}
\usepackage{pbox}
\usepackage{xcolor}

\usepackage{amsmath}
\usepackage{amssymb}
\usepackage[dvips]{graphicx}
\usepackage{epsfig}
\usepackage{algorithm}
\usepackage{algorithmic}
\usepackage{caption}
\usepackage{amsthm}
\usepackage{subcaption}
\usepackage[english]{babel}

\usepackage[ps2pdf,
bookmarks=false,
bookmarksnumbered=false, 
bookmarksopen=false, 
colorlinks=false]{}

\makeatletter

\newcommand{\Rmnum}[1]{\expandafter\@slowromancap\romannumeral #1@}
\makeatother

\newcommand{\LOS}{\text{L}}
\newcommand{\NLOS}{\text{N}}

\newcommand{\avg}{\text{avg}}

\newcommand{\EC}{\text{E}^{\text{C}}}

\newcommand{\thh}{\text{th}}
\newcommand{\st}{\text{st}}

\newcommand{\low}{\text{low}}
\newcommand{\high}{\text{high}}

\newcommand{\figsize}{.5}

\newtheorem{Theorem1}{Theorem}

\begin{document}

\title{Energy Harvesting in Unmanned Aerial Vehicle Networks with 3D Antenna Radiation Patterns}


\author{\IEEEauthorblockN{Esma Turgut, M. Cenk Gursoy, and Ismail Guvenc}
\thanks{E. Turgut and M. C. Gursoy are with the Department of Electrical
Engineering and Computer Science, Syracuse University, Syracuse, NY, 13244. I. Guvenc is with the Department of Electrical and Computer Engineering, North Carolina State University, Raleigh, NC 27695.
(e-mail: eturgut@syr.edu, mcgursoy@syr.edu, iguvenc@ncsu.edu ).}
\thanks{This work has been supported in part by the National Science Foundation (CCF-1618615). The material in this paper was presented in part at the IEEE Vehicular Technology Conference (VTC)-Fall, Honolulu, HI, Sep. 2019. \cite{Turgut_VTC}. }}

\maketitle

\begin{abstract}
In this paper, an analytical framework is provided to analyze the energy coverage performance of unmanned aerial vehicle (UAV) energy harvesting networks with clustered user equipments (UEs). Locations of UEs are modeled as a Poisson Cluster Process (PCP), and UAVs are assumed to be located at a certain height above the center of user clusters. Hence, user-centric UAV deployments are addressed. Two different models are considered for the line-of-sight (LOS) probability function to compare their effects on the network performance. Moreover, antennas with doughnut-shaped radiation patterns are employed at both UAVs and UEs, and the impact of practical 3D antenna radiation patterns on the network performance is also investigated. Initially, the path loss of each tier is statistically described by deriving the complementary cumulative distribution function and probability density function. Following this, association probabilities with each tier are determined, and energy coverage probability of the UAV network is characterized in terms of key system and network parameters for UAV deployments both at a single height level and more generally at multiple heights. Through numerical results, we have shown that cluster size and UAV height play crucial roles on the energy coverage performance. Furthermore, energy coverage probability is significantly affected by the antenna orientation and number of UAVs in the network.
\end{abstract}


\thispagestyle{empty}


\section{Introduction}
To support the unprecedented growth in demand for mobile data fueled by emerging wireless applications and the increased use of smart mobile devices, new technologies and designs are being incorporated into next-generation cellular networks. One novelty is expected to be the deployment of unmanned aerial vehicle (UAV) base stations (BSs). UAVs have been primarily
considered as high-altitude platforms at altitudes of kilometers to provide coverage in rural areas. On the other hand,
use of low-altitude UAVs has also become popular recently due to the advantage of having better link quality in shorter-distance line-of-sight (LOS) channels with the ground users. Moreover, owing to the relative flexibility in UAV deployments,
UAV BSs can be employed in a variety of scenarios including public safety communications and data collection in Internet of Things (IoT) applications \cite{Zhang2}, \cite{Zeng2}. Other scenarios include disasters, accidents, and other emergencies and also temporary events requiring
substantial network resources in the short-term such as in concerts and sporting events. In such cases, UAVs can be deployed rapidly to provide seamless wireless connectivity
\cite{Motlagh}--\cite{Bor-Yaliniz}.

In addition to growing data traffic, increasing number of devices results in a significant growth in energy demand. RF (radio frequency) energy harvesting where a harvesting device may extract energy from the incident RF signals has emerged as a promising solution to power
up low-power consuming devices \cite{Zeng}, \cite{Bi}. Therefore, the advances in energy harvesting technologies have motivated research in the study of different wireless energy harvesting networks. For example, wireless energy and/or information transfer in large-scale
millimeter-wave and microwave networks has been studied in \cite{Khan}--\cite{Renzo}. In these works, energy is harvested wirelessly from energy transmitters which are generally deployed at fixed locations. However, low-power consuming devices can potentially be distributed in a large area. In such cases, the performance of energy harvesting will be limited by the low end-to-end power transmission efficiency due to the loss of RF signals over long distances \cite{Xu}.

In order to improve the efficiency, instead of fixed energy transmitters such as ground base stations (BSs), the deployment of mobile energy transmitters is proposed recently. In particular, although the UAVs are typically power-limited,  UAV-assisted energy harvesting has become attractive due to the flexibility and relative ease in deploying UAV BSs. In \cite{Wu}, mobility of the UAV with a directional antenna is exploited by jointly optimizing the altitude, trajectory, and transmit beamwidth of the UAV in order to maximize the energy transferred to two energy receivers over a finite charging period. In \cite{Xu}, authors consider a more general scenario with more than two energy receivers where the amount of received energy by all energy receivers is maximized via trajectory control. In \cite{Hu}, a UAV-enabled wireless power transfer network is studied as well. Minimum received energy among all ground nodes is maximized by optimizing the UAV's one-dimensional trajectory. Both downlink
wireless power transfer and uplink information transfer is considered in \cite{Xie} with one UAV and a set of ground users. The UAV charges the users in downlink and users use the harvested energy to send the information to the UAV in the uplink. Similarly, a wireless-powered communication network with a mobile hybrid access point UAV is considered in \cite{Cho}. In this work, the UAV performs weighted energy transfer and receives information from the far-apart nodes based on the weighted harvest-then-transmit protocol. The use of UAVs to power up energy constrained sensor nodes has been suggested in \cite{Caillouet}. UAVs are considered to be equipped with the dedicated chargers facing the ground so that they can recharge the sensors' batteries using RF energy harvesting. A practical example is AT\&T's flying Cell on Wings (COW) drones \cite{ATT}. In particular, AT\&T's COW is a UAV equipped with a small base station. Flying COW is tethered to a power source on the ground with a cable. In these types of practical implementations, power constraints are more relaxed and such a UAV can provide both cell service and also supply wireless energy to sensors and small IoT devices.

In a separate line of research in the literature, the performance of UAV-assisted wireless networks is extensively studied recently. Similar to 2D networks, stochastic geometry has been employed in the network level analysis of UAV networks by considering UAVs distributed randomly in 3D space. Effect of different network parameters on the coverage probability is explored in several recent works such as \cite{Galkin}--\cite{Zhou}. In \cite{Liu}, authors analyzed the network performance for three different type of LOS probability models. Spectrum sharing in UAV networks is analyzed in \cite{Zhang}--\cite{Lyu}. Additionally, optimal deployment of UAVs is investigated in \cite{Bor-Yaliniz2}--\cite{Mozaffari1}.

It is important to note that the antenna number, type, and orientation are critical factors that affect the performance in UAV-assisted
networks. Indeed, several recent studies, e.g., \cite{Chandhar} and \cite{Geraci}, have addressed scenarios in which antenna arrays are deployed in UAV-assisted cellular networks. Similarly, in \cite{Lyu2}, the authors have considered directional antennas for UAVs. However, a practical antenna pattern which is omnidirectional in the horizontal plane but directional in the vertical plane is employed for ground BSs. Regarding the antenna type, omnidirectional antennas
can also be used in UAVs, especially considering their mobility \cite{Khawaja}. At the same time, since even the UAV's own body can shadow the antenna and result in a poor link quality, the orientation of the antennas plays
an important role on the performance \cite{Yanmaz}. There has been limited analytical and experimental works
studying the effect of three dimensional (3D) antenna radiation patterns on the link quality between the
UAV and ground users. In \cite{Yanmaz}, impact of antenna orientation is investigated by placing two antennas
on a fixed wing UAV flying on a linear path with 802.11a interface. Similarly, path loss and small-scale fading
characteristics of UAV-to-ground user links are analyzed with a simple antenna extension to 802.11 devices in
\cite{Yanmaz2}. In \cite{Chen}, ultra-wideband (UWB) antennas with doughnut-shaped radiation patterns are employed at both UAVs
and ground users to analyze the link quality at different link distances, UAV heights, and antenna orientations. The large bandwidth of UWB radio signals is utilized in the measurements to obtain a high temporal resolution of multipath components. Authors develop a simple analytical model to approximate the impact of the 3D antenna radiation pattern on the received
signal. However, none of these works study the effect of UAV antenna orientation on the network performance.

\subsection{Contributions and Organization}
In this paper, we consider a UAV network consisting of UAVs operating at a certain altitude above ground. The locations of user equipments (UEs) are modeled as a Poisson cluster process (PCP), and the UAVs are assumed to be located at a certain height above the center of user clusters. Since UAVs are deployed in overloaded scenarios, locations of UAVs and UEs are expected to be correlated
and UEs are more likely to form clusters. Hence, modeling the UE locations by PCP is more appropriate and realistic.
Moreover, we consider that UWB antennas with doughnut-shaped radiation patterns are employed at both UAVs and UEs, and we study the effect of practical 3D antenna radiation patterns on energy harvesting from UAVs.

More specifically, our main contributions can be summarized as follows:
\begin{itemize}
\item An analytical framework is provided to analyze energy coverage performance of a UAV network with clustered UEs. UE locations are assumed to be PCP distributed to capture the correlations between
the UAV and UE locations.

\item We divide the network into two tiers: $0^{\thh}$ tier UAV and $1^{\st}$ tier UAVs. $0^{\thh}$ tier UAV is the
cluster center UAV around which the typical UE is located, while other UAVs constitute the $1^{\st}$ tier.

\item Two different LOS probability functions, i.e., a high-altitude model and a low-altitude model, are considered in order
to investigate and compare their impact on the network performance.

\item Different from the previous studies, more practical antennas with
 doughnut-shaped radiation patterns are employed at both UAVs and UEs to provide a more realistic performance
 evaluation for the network.

\item We first derive the complementary cumulative distribution functions (CCDFs) and the probability density functions (PDFs) of the path losses for each tier, then obtain the association probabilities
by using the averaged received power UAV association rule.

\item Average harvested power expression is obtained. Then, total energy coverage probability is determined by deriving the Laplace transforms of the interference terms arising from each tier. Energy coverage is characterized for UAV deployments at a single height level and also at multiple heights.

\end{itemize}

The rest of the paper is organized as follows. In Section \ref{sec:system_model}, system model is introduced, and path loss, blockage, and 3D antenna models are described.
Path loss and association probabilities are statistically characterized in Section \ref{sec:Path Loss and Cell Association}. In Section
\ref{sec:Energy Coverage_Probability Analysis}, energy coverage probability of the UAV network is determined. In Section
\ref{sec:Simulation and Numerical Results}, simulations and numerical results are provided, demonstrating the impact of several key
parameters on the energy coverage performance of the network. Finally, Section
\ref{sec:Conclusion} concludes the paper. Proofs are included in the Appendix.

\section{System Model} \label{sec:system_model}
In this section, we describe the system model of the UAV network with clustered UEs. We address a downlink network, in which the spatial distribution of the UAVs is modeled by an independent homogeneous PPP $\Phi_{U}$ with density $\lambda_{U}$ on the Euclidean plane.
The height of UAVs is denoted by $H$. Note that UAVs can be used to offload traffic from the ground cellular BSs and reduce congestion around hotspots. In energy harvesting applications, UAVs can be used to transfer energy to e.g., ground sensors and low-power IoT devices, to  energize them. They can also be deployed in case of emergencies during which ground infrastructure is strained \cite{Guvenc}.
In our model,  the UEs are clustered around the projections of UAVs on the ground, and the locations of the clustered UEs is  described by a PCP, denoted
by $\Phi_C$. In applications involving UAVs, UEs are expected to be located in high UE density areas, forming clusters. Consequently, modeling of UE distribution as a PCP rather than a homogeneous PPP is more accurate.

\begin{figure}
\centering
  \includegraphics[width=\figsize\textwidth]{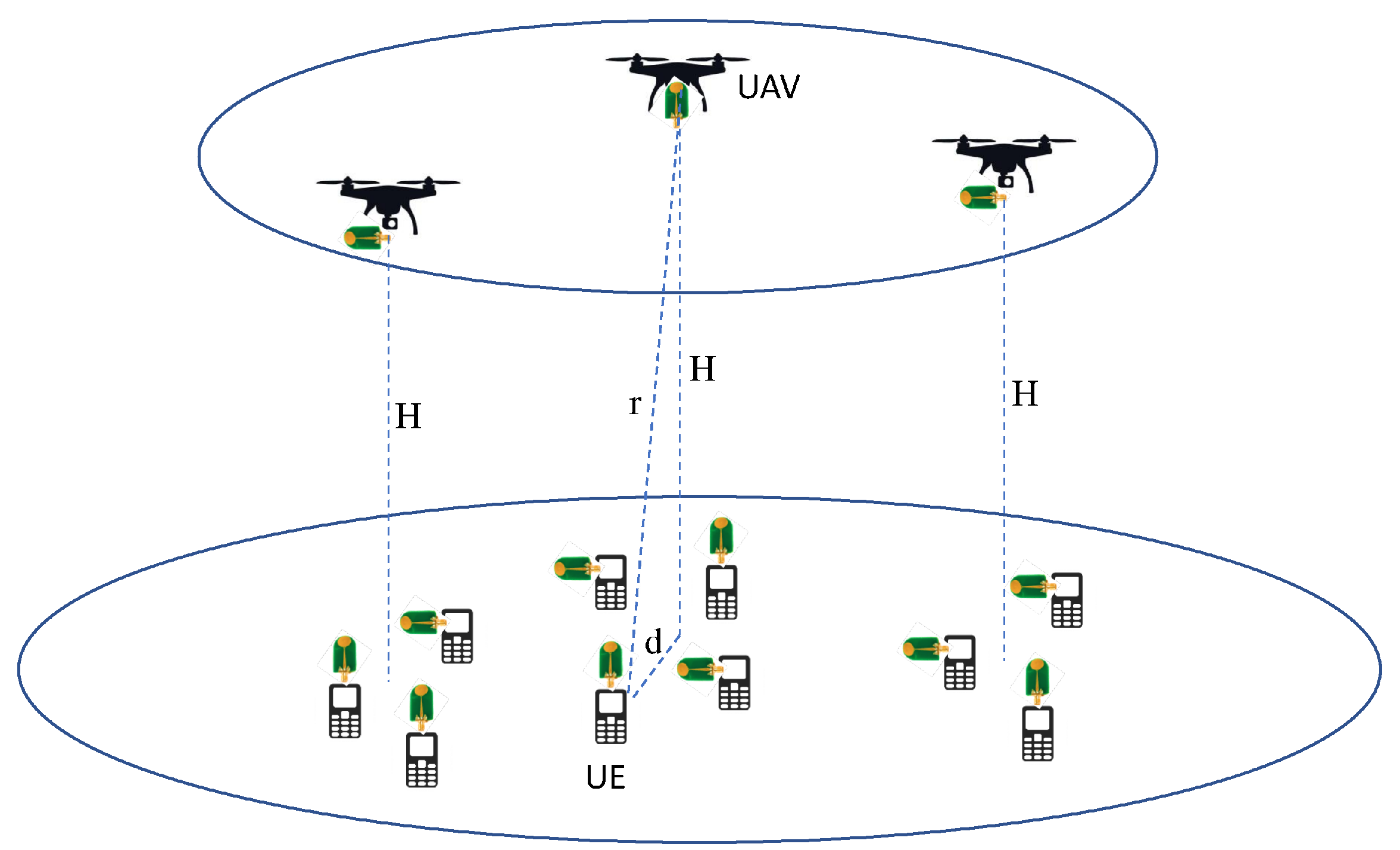}
  \caption{\small Network model for a UAV energy harvesting network. BSs are distributed as a PPP, while UEs are Gaussian distributed around the cluster centers (projections of UAVs on the ground). Both BS and UEs are equipped with UWB antennas with different antenna orientations.    \normalsize}
\label{Fig_Network_Model}
\end{figure}

In this paper, we model $\Phi_C$ as a Thomas cluster process, where the UEs are symmetrically independently and identically distributed (i.i.d.) around the cluster centers (which are projections of UAVs on the ground), according to a Gaussian distribution with zero mean and variance $\sigma_c^2$. Therefore, the UE's location is statistically described by the following PDF and CCDF \cite{Haenggi}:
\begin{align}
f_{D}(d)&=\frac{d}{\sigma_c^2} \exp\left( -\frac{d^2 }{2\sigma_c^2}\right), \quad d \geq 0, \label{PDF_of_d} \\
\bar{F}_{D}(d)&=\exp\left( -\frac{d^2 }{2\sigma_c^2}\right), \quad d \geq 0, \label{Fbar_R}
\end{align}
respectively, where $d$ is the 2D distance of a UE with respect to the cluster center on the ground. Without loss of generality, we perform the analysis for a typical UE which is randomly chosen from a randomly chosen cluster. Since cluster centers; i.e. UAVs, are PPP distributed and the PPP is stationary, location of this typical UE can be transformed to the origin according to Slivnyak's theorem \cite{Saha}. The typical UE is assumed to be is associated with the UAV providing the maximum average received power. Although we have only a single-tier network composed of UAVs, we also consider an additional tier, named as $0^{\thh}$ tier that only includes the cluster center of the typical UE similarly as in \cite{Esma} and \cite{Saha}. Therefore, overall UAV density is equal to tier 1 density, and tier 0 has only one UAV, which is the cluster center of the typical UE. Essentially tier 0 is introduced to differentiate the cluster-center UAV from other UAVs, because the distance distribution equations of the typical UE to its own cluster-center UAV given in (\ref{PDF_of_d}) and (\ref{Fbar_R}) are different from the distribution of the distances to other UAVs. Thus, our network model can be considered as a two-tier network consisting of a $0^{\thh}$ tier cluster-center UAV and $1^{\st}$ tier UAVs. The considered network model is depicted in Fig. \ref{Fig_Network_Model}.


\subsection{Path Loss and Blockage Modeling}
A transmitting UAV can either have a line-of-sight (LOS) or non-line-of-sight (NLOS) link to the typical UE. Consider an arbitrary
link of length $r$ between a UE and a UAV, and define the LOS probability function as the probability that the
link is LOS. Different LOS probability functions have been used in the literature.
In this paper, we consider the two models proposed in \cite{Al-Hourani} and \cite{3GPP}, which are high-altitude and low-altitude models, respectively.

High-altitude model is widely used especially in satellite communications where the altitude is around hundred of kilometers.
It has also been widely employed in UAV-assisted networks recently. LOS probability function for the high-altitude model is given as follows:
\begin{equation}
\mathcal{P}_{\LOS}^{\high}(r)=\left(\frac{1}{1+b\exp\left(-c\left(\frac{180}{\pi}\sin^{-1}\left(\frac{H}{r}\right)-b\right)\right)}\right), \label{LOS_probability1}
\end{equation}
where $r$ is the 3D distance between the UE and UAV, $H$ is the UAV height, $b$ and $c$ are constants whose values depend on the environment.
It can be easily verified that the LOS probability in (\ref{LOS_probability1}) increases as the UAV height, $H$, increases.

\begin{figure}
    \centering
    \begin{subfigure}[b]{0.45\textwidth}
        \includegraphics[width=\textwidth]{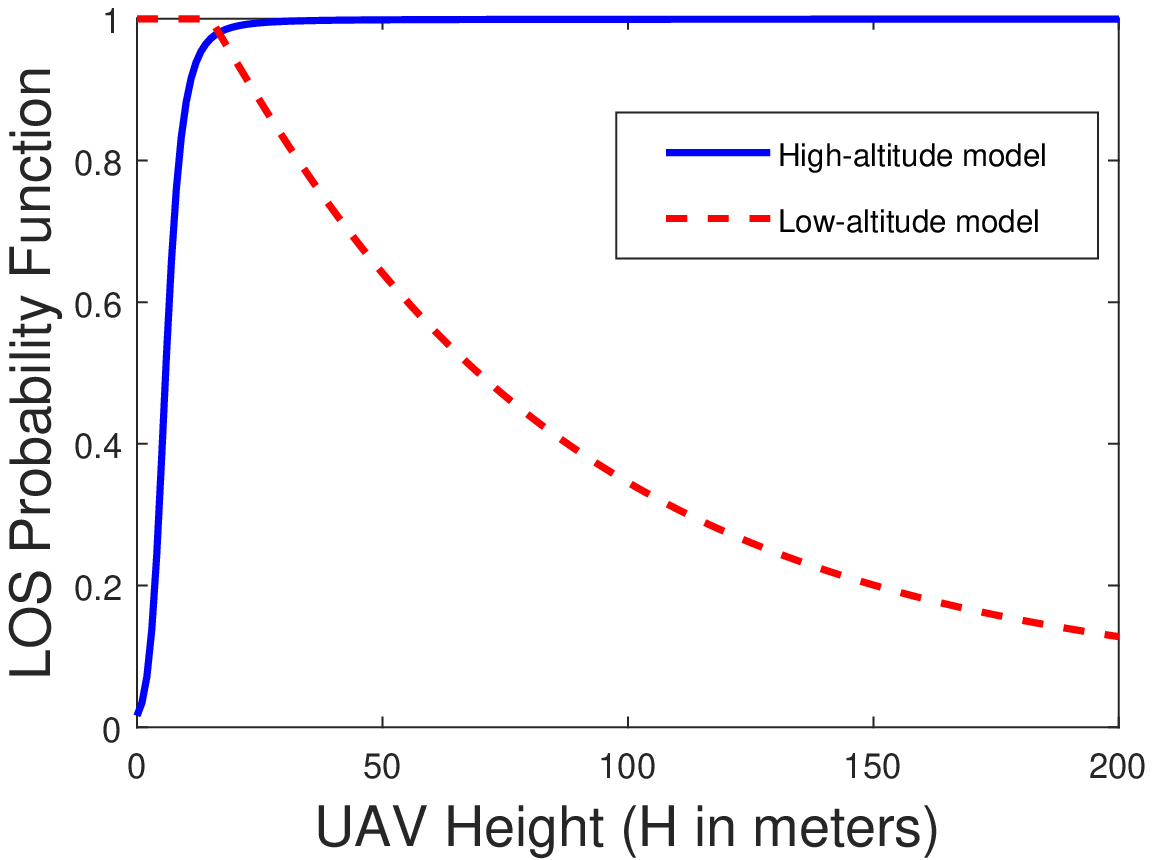}
        \caption{}
        \label{Fig_LOS_funcH}
    \end{subfigure}
    ~
    \begin{subfigure}[b]{0.45\textwidth}
        \includegraphics[width=\textwidth]{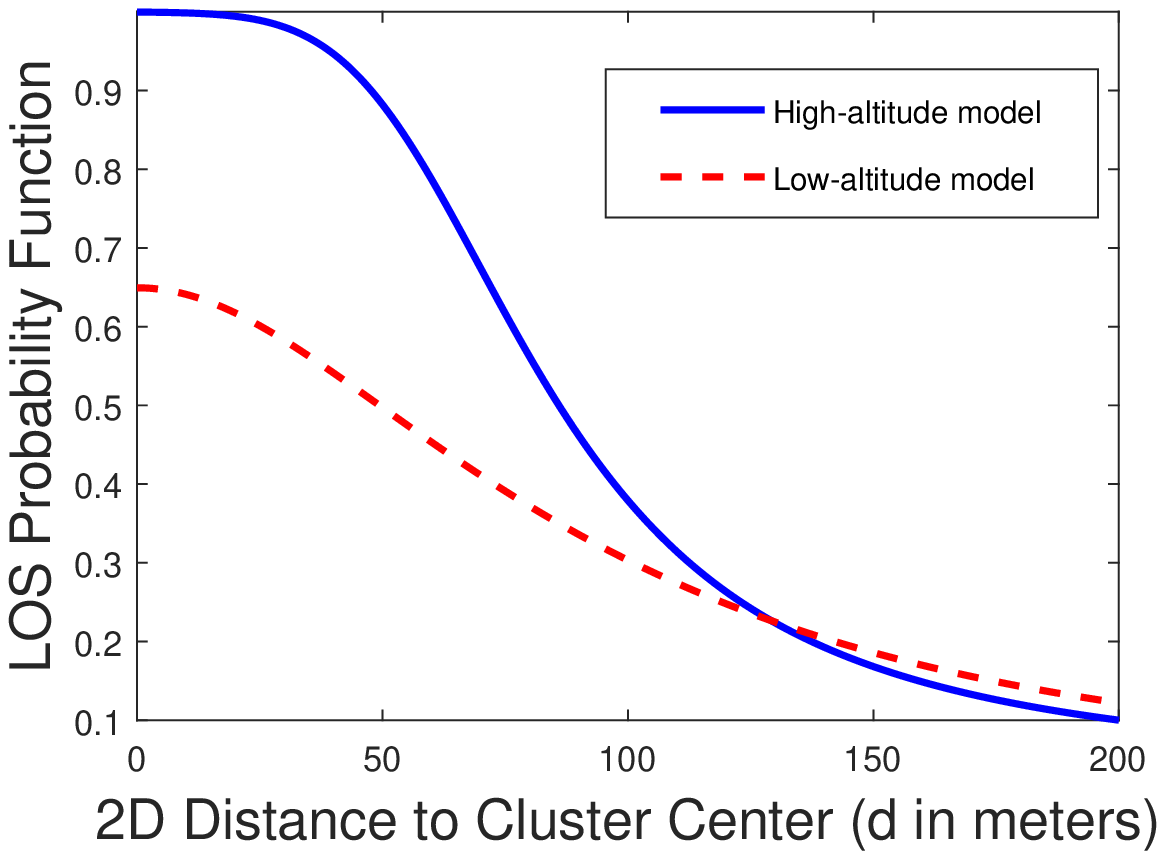}
        \caption{}
        \label{Fig_LOS_funcd}
    \end{subfigure}
    \caption{\small LOS probability function for high-altitude and low-altitude models as a function of (a) UAV height $H$ and (b) 2D distance to the UAV $d$.  \normalsize}\label{Fig_LOS_func}
\end{figure}

Since practical values for UAV height in certain applications is around $50\sim100$ meters, a more realistic LOS probability function proposed for 3GPP terrestrial
communications is employed also for UAV networks in \cite{Liu}. The height of a macrocell base station is usually around 32 m, which is
comparable to the practical UAV height. Therefore, employment of the LOS probability function for 3GPP macrocell-to-UE communciation is
also reasonable for the UAV networks in such relatively low-altitude scenarios. For the low-altitude model, LOS probability function is expressed as
\begin{equation}
\mathcal{P}_{\LOS}^{\low}(r)= \min\left(1,\frac{18}{r}\right)\left(1-\exp\left(-\frac{r}{63}\right)\right)+\exp\left(-\frac{r}{63}\right). \label{LOS_probability2}
\end{equation}
Note that different from the high-altitude model, LOS probability function in (\ref{LOS_probability2}) decreases with the increase in the 3D distance $r$, independent of the
UAV height. In Fig. \ref{Fig_LOS_func}, LOS probability function is plotted using high-altitude and low-altitude models.
Solid lines show the LOS probability as a function of the UAV height $H$ when the 2D distance to the UAV is fixed at $d=10$ m, and
dashed lines display the LOS probability as a function of the 2D distance to the UAV $d$ when the UAV height is $H=50$ m.
As shown in Fig. \ref{Fig_LOS_func}, LOS probability increases with increasing UAV height when the high-altitude model is used, and decreases when the low-altitude model is considered. We observe that the LOS probability decreases for both models as the 2D distance to the UAV increases. We also note that the analysis in the remainder of the paper is general and is applicable to any LOS probability function. Only in Section \ref{sec:Simulation and Numerical Results}, we employ the LOS probability functions in (\ref{LOS_probability1}) and (\ref{LOS_probability2}) to obtain the numerical results.

In general, NLOS links experience higher path loss than the LOS links due to signals being reflected and scattered in NLOS propagation environments. Therefore,
different path loss laws are applied to LOS and NLOS links. Thus, the path loss on each link in tier $k \in \{0,1\}$ can be expressed
as
\begin{equation}\label{PL_model}
\begin{split}
L_{k,\LOS}(r)&= r^{\alpha_{\LOS}}  \\
L_{k,\NLOS}(r)&= r^{\alpha_{\NLOS}} ,
\end{split}
\end{equation}
where $\alpha_{\LOS}$ and $\alpha_{\NLOS}$ are the LOS and NLOS path-loss exponents, respectively.

\subsection{3D Antenna Modeling}
\begin{figure}
\centering
  \includegraphics[width=\figsize\textwidth]{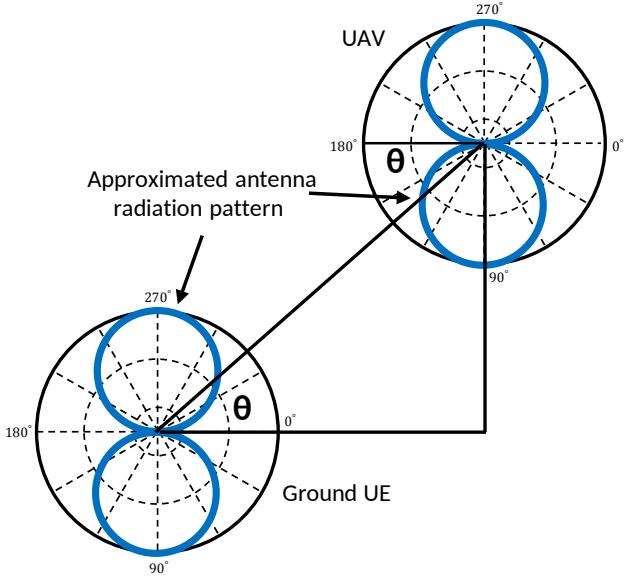}
  \caption{\small Approximated antenna radiation pattern for HH antenna orientation.  \normalsize}
\label{Fig_Antenna_Model}
\end{figure}

In this paper, we consider the analytical model developed in \cite{Chen} for the effect of 3D antenna radiation patterns on the received
signal. UWB transmitter and receiver antennas with doughnut-shaped radiation patterns centered at a frequency of 4 GHz are placed at the UAV and UE, respectively,
and air-to-ground channel measurements are carried out in order to characterize the impact of the 3D antenna radiation pattern on the
received signal for different antenna orientations in \cite{Chen}. As a result of these measurements, transmitter and receiver antenna
gains are modeled analytically for horizontal-horizontal (HH), horizontal-vertical (HV) and vertical-vertical (VV) antenna orientations
as follows:
\begin{equation}
G_k(\theta)=G_{\text{TX}}(\theta)G_{\text{RX}}(\theta)=\begin{cases}
\sin(\theta)\sin(\theta) & \text{for} \quad \text{HH} \\
\sin(\theta)\cos(\theta) & \text{for} \quad \text{HV} , \\
\cos(\theta)\cos(\theta) & \text{for} \quad \text{VV} \label{Antenna_gain}
\end{cases}
\end{equation}
where $\theta$ is the elevation angle between the transmitter at the UAV and the receiver at the UE on the ground. In this antenna model,
radiation pattern is approximated by a circle in the vertical dimension, while it is assumed to be constant for all horizontal
directions. In other words, antenna gains depend only on the elevation angle $\theta$, and are considered as independent of the azimuth angle
between the transmitter at the UAV and the receiver at the UE. Approximated antenna radiation patterns of UAV and UE are shown in Fig.
\ref{Fig_Antenna_Model} for HH antenna orientation.
They can be plotted for HV and VV orientations as well by rotating the transmitter and/or receiver antennas by $90^{\circ}$. Note that
for HH antenna orientation $G_{\text{TX}}(\theta)=G_{\text{RX}} \rightarrow 0$ as $\theta \rightarrow 0$ which happens when the UEs
are located far away from the cluster center, i.e. as the $\sigma_c$ increases, and $G_{\text{TX}}(\theta)=G_{\text{RX}} \rightarrow 1$
as $\theta \rightarrow 90^{o}$ which happens when the UEs get closer to the cluster center. Similar observations can be drawn for VH and
VV antenna orientations. Effective antenna gain $G_k$ as a function of $r$ can be rewritten in terms of UAV height $H$ and
the path loss on each link in tier $k \in \{0,1\}$ as
\begin{equation}\small
G_k(r)=\begin{cases}
H^2 L_{k,s}^{-\frac{2}{\alpha_s}}(r) & \text{for} \quad \text{HH} \\
H \left(\sqrt{L_{k,s}^{\frac{2}{\alpha_s}}(r)-H^2} \right) L_{k,s}^{-\frac{2}{\alpha_s}}(r) & \text{for} \quad \text{HV} \\
 \left(L_{k,s}^{\frac{2}{\alpha_s}}(r)-H^2 \right) L_{k,s}^{-\frac{2}{\alpha_s}}(r) & \text{for} \quad \text{VV}.
\end{cases} \normalsize
\end{equation}

In the rest of the analysis, we assume that the typical UE and all UAVs in the network have horizontal antenna orientation. Therefore, HH antenna orientation for the main link and interfering links are considered due to its analytical tractability. Moreover, UEs are considered to be clustered around the projections of UAVs on the ground and more UEs are encouraged to be associated with their cluster center UAV. As a result, the angle between the transmitter at the UAV and the receiver at the UE is expected to be large. Therefore, HH antenna orientation is more suitable than the other two orientations. However, in the numerical results section, simulation results
for HV and VV orientations are also provided in order to compare their effect on the UAV association and energy coverage probabilities.

Finally, we note that a summary of notations is provided in Table \ref{Table_notation} below.

\begin{table}[htbp]
	\caption{Table of Notations}
	\centering
	\begin{tabular}{|l|p{2.5in}|}
		\hline
		\footnotesize \textbf{Notations} &  \footnotesize  \textbf{Description}  \\ \hline
		
		\scriptsize$\Phi_U, \lambda_U, \Phi_C $&  \scriptsize PPP of UAVs, the density of $\Phi_U$, PCP of UEs.  \\ \hline
        \scriptsize$H$  & \scriptsize Height of UAVs.\\\hline
		\scriptsize$\sigma_c$  & \scriptsize  UE distribution's standard deviation.\\\hline
        \scriptsize$d$  & \scriptsize  2D distance of a UE with respect to the cluster center on the ground.\\\hline
        \scriptsize$r$  & \scriptsize  3D distance between the UE and UAV.\\\hline
        \scriptsize$\mathcal{P}_{\LOS}^{\high}(r), \mathcal{P}_{\LOS}^{\low}(r)$ &\scriptsize LOS probability functions for high and low altitude models.          \\ \hline
        \scriptsize$b, c$ &\scriptsize Environment dependent constants.          \\ \hline
        \scriptsize$L_{k,\LOS}(r), L_{k,\NLOS}(r)$ &\scriptsize Path losses on LOS and NLOS links in the $k^{th}$ tier.          \\ \hline
        \scriptsize$\alpha_{\LOS}, \alpha_{\NLOS}$ &\scriptsize LOS and NLOS path-loss exponents.          \\ \hline
		\scriptsize$G_{\text{TX}}(\theta), G_{\text{RX}}(\theta)$  &\scriptsize  Transmitter and receiver antenna gains. \\ \hline
        \scriptsize$G_k(r)$  &\scriptsize  Effective antenna gain. \\ \hline
        \scriptsize$\theta$  &\scriptsize  Elevation angle between the transmitter and the receiver. \\ \hline
        \scriptsize$\bar{F}_{L_{k}}(x)$  &\scriptsize  CCDF of the path loss from a typical UE to a $k^{th}$ tier UAV. \\ \hline
        \scriptsize$f_{L_{k,s}}(x)$  &\scriptsize  PDF of the path loss from a typical UE to a $k^{th}$ tier LOS/NLOS UAV. \\ \hline
    	\scriptsize$P_k$  &\scriptsize  Transmit power of UAVs in the $k^{th}$ tier. \\ \hline
    	\scriptsize$h$   &\scriptsize Small-scale fading gain. \\ \hline	
		\scriptsize $\mathcal{A}_{k,s}$  & \scriptsize Association probability with a $k^{th}$ tier LOS/NLOS UAV.  \\ \hline
		\scriptsize $I_{j,k}$   &   \scriptsize Interference from the $j^{th}$ tier UAV, when the main link is in the $k^{th}$ tier.  \\ \hline
		\scriptsize $\mathcal{E}_{k,0}$   &   \scriptsize Exclusion disc of path loss, inside which no interference exists.  \\ \hline
		\scriptsize $P_{r,k}$   &   \scriptsize Total received power  \\ \hline
   		\scriptsize $\EC_{k,s}(\Gamma_k)$   &   \scriptsize Conditional energy coverage probability given that the UE is associated with a $k^{\thh}$ tier LOS/NLOS UAV.    \\ \hline	
   		\scriptsize $\Gamma_k$   &   \scriptsize Energy outage threshold.  \\ \hline	
   		\scriptsize$\xi$ &\scriptsize Rectifier efficiency.          \\ \hline
		\scriptsize $\mathcal{L}_{I_{j,k}}(\Gamma_k,\mathcal{E}_{k,0})$   &   \scriptsize Laplace transform of $I_{j,k}$. \\ \hline	
	\end{tabular} \label{Table_notation}
\end{table}\normalsize

\section{Path Loss and UAV Association} \label{sec:Path Loss and Cell Association}
\subsection{Statistics of the Path Loss}
In this section, we provide statistical characterizations by identifying the CCDF and the PDF of the path loss.

\emph{Lemma 1:} The CCDF of the path loss from a typical UE to a $0^{\thh}$ tier UAV can be formulated as
\begin{align}
 \bar{F}_{L_{0}}(x)&= \hspace{-0.3cm} \sum_{s \in \{\LOS,\NLOS\}}  \bar{F}_{L_{0,s}}(x) \nonumber \\
&= \hspace{-0.3cm} \sum_{s \in \{\LOS,\NLOS\}} \int_{\sqrt{x^{2/\alpha_{s}}-H^2}}^{\infty} \mathcal{P}_s(\sqrt{d^2+H^2}) f_D(d) \mathrm{d}d,  \label{CCDF_0}
\end{align}
where $f_D(d)$ is given in (\ref{PDF_of_d}), and $\mathcal{P}_{s}(\cdot)$ is the LOS or NLOS probability depending on whether $s = \LOS$ or $s = \NLOS$ \footnote{For instance, LOS probability is given by (\ref{LOS_probability1}) and (\ref{LOS_probability2}) for the high-altitude and low-altitude models, respectively, and NLOS probability is $\mathcal{P}_{\NLOS}(\cdot) = 1-\mathcal{P}_{\LOS}(\cdot)$.}.

\emph{Proof:} See Appendix \ref{Proof of Lemma 1}. \hfill $\square$

\emph{Lemma 2:} CCDF of the path loss from a typical UE to a $1^{\st}$ tier UAV is
\begin{equation}
\bar{F}_{L_1}(x)= \prod_{s \in \{\LOS,\NLOS\}} \bar{F}_{L_{1,s}}(x) = \prod_{s \in \{\LOS,\NLOS\}} \exp\big(-\Lambda_{1,s}([0,x))\big), \label{CCDF_1}
\end{equation}
where $\Lambda_{1,s}([0,x))$ is given by
\begin{align}
\Lambda_{1,s}([0,x))= 2\pi\lambda_U \int_{H}^{x^{\frac{1}{\alpha_{s}}}} \mathcal{P}_{s}(r) r \mathrm{d}r . \label{intensity_function_1}
\end{align}

\emph{Proof:} See Appendix \ref{Proof of Lemma 2}. \hfill $\square$

\emph{Corollary 1:} The PDF of the path loss from a typical UE to a $0^{\thh}$ tier LOS/NLOS UAV can be determined by applying the Leibniz integral rule as follows:
\begin{align}
f_{L_{0,s}}(x)&=-\frac{d\bar{F}_{L_{0,s}}(x)}{dx}   \nonumber \\
              & =\frac{1}{\sigma_c^2} \frac{x^{\frac{2}{\alpha_{s}}-1}}{\alpha_s} \mathcal{P}_{s}\left(x^{\frac{1}{\alpha_{s}}}\right) \exp\left(-\frac{1}{2\sigma_c^2} \left(x^{\frac{2}{\alpha_{s}}}-H^2\right)\right). \label{f_L0s}
\end{align}

\emph{Corollary 2:} The PDF of the path loss from a typical UE to a $1^{\st}$ tier LOS/NLOS UAV is
\begin{equation}
f_{L_{1,s}}(x)=-\frac{d\bar{F}_{L_{1,s}}(x)}{dx}=\Lambda_{1,s}^{\prime}([0,x)) \exp\big(-\Lambda_{1,s}([0,x))\big),   \label{f_L1s}
\end{equation}
where $\Lambda_{1,s}^{\prime}([0,x))$ is given by
\begin{equation}
\Lambda_{1,s}^{\prime}([0,x)) =2\pi\lambda_U \frac{x^{\frac{2}{\alpha_{s}}-1}}{\alpha_{s}}\mathcal{P}_{s}\left(x^{\frac{1}{\alpha_{s}}}\right) \label{Lambda_1s_prime}
\end{equation}
by again applying the Leibniz integral rule.

In the results above, we have determined the CCDFs and PDFs of the path loss for each tier. They depend on the key network parameters including the variance of the cluster process $\sigma_c^2$, UAV density $\lambda_U$, UAV LOS probability $\mathcal{P}_{s}(\cdot)$, UAV height $H$ and path loss exponents $\alpha_s$. In the following sections, these distributions are utilized in determining the association and energy coverage probabilities.

\subsection{UAV Association}
In this work, we assume that the UEs are associated with a UAV providing the strongest long-term averaged power. In other words, a typical UE is associated with its cluster center UAV, i.e., the $0^{\thh}$ tier UAV, if
\begin{equation}
P_0 G_0(r) L_0^{-1}(r) \geq P_1 G_1(r) L_{\text{min},1}^{-1}(r),
\end{equation}
where $P_k$ and $G_k(r)$ denote the transmit power and antenna gain of the link, respectively, in tier $k \in (0,1)$. $L_0(r)$ is the path loss from the $0^{\thh}$ tier UAV, and $L_{\text{min},1}(r)$ is the path loss from $1^{\st}$ tier UAV providing the minimum path loss. In the following lemma, we provide the association probabilities using the result of Lemmas 1, Lemma 2, Corollary 1 and Corollary 2.

\emph{Lemma 3:} The association probabilities with a $0^{\thh}$ tier LOS/NLOS UAV and  $1^{\st}$ tier LOS/NLOS UAV are given, respectively, as
\begin{align}
\mathcal{A}_{0,s} = \int_{ H^{\alpha_{s}}}^{\infty} &\prod_{m \in \{\LOS,\NLOS\}} \bar{F}_{L_{1,m}}\left(\left(\frac{P_1}{P_0}l_{0,s}^{\frac{2}{\alpha_s}+1}\right)^{\frac{\alpha_m}{\alpha_m+2}}\right) \nonumber \\
        &   \times f_{L_{0,s}}(l_{0,s}) \mathrm{d}l_{0,s},  \label{Association_Prob0}
\end{align}
\begin{align}
\mathcal{A}_{1,s}=  \int_{ H^{\alpha_{s}}}^{\infty} & \sum_{m \in \{\LOS,\NLOS\}} \bar{F}_{L_{0,m}}\left(\left(\frac{P_0}{P_1}l_{1,s}^{\frac{2}{\alpha_s}+1}\right)^{\frac{\alpha_m}{\alpha_m+2}}\right) \nonumber \\
& \times \bar{F}_{L_{1,s^{\prime}}}(l_{1,s}) f_{L_{1,s}}(l_{1,s})\mathrm{d}l_{1,s},  \label{Association_Prob1}
\end{align}
where $s, s' \in \{\LOS,\NLOS\}$ and $s \neq s'$.

\textit{Proof}: See Appendix \ref{Proof of Lemma 3}. \hfill $\square$

In the corollary below, we provide a closed-form expression for the association probability in a special case with which the effects of different parameters on association probability can be easily identified.

\emph{Corollary 3:} Consider the same UAV network with $P_0=P_1$ and the LOS probability $\mathcal{P}_{\LOS}(\cdot)=1$, i.e., all UAVs are LOS to the typical UE. Then, the association probabilities specialize to the following expressions (which also confirm the results in \cite{Wang}):
\begin{align}
\mathcal{A}_{0,L}&=\frac{1}{1+2\pi\lambda_U\sigma_c^2} \\
\mathcal{A}_{1,L}&=\frac{2\pi\lambda_U\sigma_c^2}{1+2\pi\lambda_U\sigma_c^2}
\end{align}
with $\mathcal{A}_{0,N}=\mathcal{A}_{1,N}=0$. According to the these results, a typical UE obviously prefers to connect to the $0^{\thh}$ tier UAV when the value of $\sigma_c$ is small, and connect to a $1^{\st}$ tier UAV for higher values of $\sigma_c$ and $\lambda_U$.

\section{Energy Coverage Probability Analysis} \label{sec:Energy Coverage_Probability Analysis}
In this section, we use stochastic geometry and adopt an analytical framework to characterize the energy coverage probability for a typical UE clustered around the $0^{\thh}$ tier UAV (i.e., its own cluster-center UAV).

\subsection{Downlink Power Transfer}
The total power received at a typical UE at a random distance $r$ from its associated UAV in the $k^{\thh}$ tier can be written as
\begin{align}
P_{r,k}&=S_k+ \sum_{j=0}^{1} I_{j,k} \quad \text{for} \quad k=0,1, \label{total_received_power}
\end{align}
where the received power from the serving UAV $S_k$ and the interference power received from the UAVs in the $j^{\text{th}}$ tier $I_{j,k}$ are given as follows:
\begin{align}
S_k&=P_k G_k(r) h_{k,0} L_k^{-1}(r), \label{received_power} \\
I_{0,1}&= P_0 G_0(r)h_{0,0} L_0^{-1}(r), \label{interference0}\\
I_{1,k}&= \sum_{i \in \Phi_{U}\setminus{\mathcal{E}_{k,0}}} P_1 G_{i}(r) h_{1,i} L_{i}^{-1}(r), \label{interference1}
\end{align}
where $h_{k,0}$ and $h_{j,i} $ are the small-scale fading gains from the serving and interfering UAVs, respectively.
Note that since only one UAV exists in the $0^{\thh}$ tier, $I_{0,0}=0$. $h$ denotes the small-scale fading gain and is assumed to be exponentially distributed. From the UAV association policy,
when a typical UE is associated with a UAV whose path loss is $L_k(r)$, there exists no UAV within a disc $\mathcal{E}_{k,0}$
centered at the origin, which is also known as the exclusion disc. In this work, we also consider a linear energy harvesting model in which energy can be harvested if the received power is larger than zero. Therefore, the average harvested power at a typical UE is given in the following theorem.

\emph{Lemma 4:} The average harvested power at a typical UE at a random distance $r$ from its associated UAV in the $k^{\thh}$ tier is given at the top of the next page in (\ref{lemma4})
\begin{figure*}
\begin{align}
P^{\avg}&= \sum_{s \in \{\LOS,\NLOS\}} \Bigg[ \int_{H^{\alpha_{s}}}^{\infty}
\left[P_0H^2 l_{0,s}^{-\left(1+\frac{2}{\alpha_s}\right)}+\Psi_{I_{1,0}}(\mathcal{E}_{0,0})\right]  \prod_{m \in \{\LOS,\NLOS\}} \bar{F}_{L_{1,m}}\left(\left(\frac{P_1}{P_0}l_{0,s}^{\frac{2}{\alpha_s}+1}\right)^{\frac{\alpha_m}{\alpha_m+2}}\right) f_{L_{0,s}}(l_{0,s}) \mathrm{d}l_{0,s}  \nonumber \\
&+\int_{H^{\alpha_{s}}}^{\infty}
\left[P_1H^2 l_{1,s}^{-\left(1+\frac{2}{\alpha_s}\right)}+\sum_{j=0}^{1}\Psi_{I_{j,1}}(\mathcal{E}_{1,0})\right]  \sum_{m \in \{\LOS,\NLOS\}} \bar{F}_{L_{0,m}}\left(\left(\frac{P_0}{P_1}l_{1,s}^{\frac{2}{\alpha_s}+1}\right)^{\frac{\alpha_m}{\alpha_m+2}}\right) \bar{F}_{L_{1,s^{\prime}}}(l_{1,s}) f_{L_{1,s}}(l_{1,s}) \mathrm{d}l_{1,s}\Bigg] \label{lemma4}
\end{align}
\end{figure*}
where
\begin{align}
\Psi_{I_{0,k}}(\mathcal{E}_{k,0}) = \sum_{s^{\prime} \in \{\LOS,\NLOS\}} \int_{\mathcal{E}_{k,0}}^{\infty} P_0 H^2 x^{-\left(1+\frac{2}{\alpha_{s^{\prime}}}\right)}  f_{L_{0,s^{\prime}}}(x) \mathrm{d}x, \label{Psi0}
\end{align}
\begin{align}
\Psi_{I_{1,k}}(\mathcal{E}_{k,0})= \sum_{s^{\prime} \in \{\LOS,\NLOS\}} \int_{\mathcal{E}_{k,0}}^{\infty} P_1 H^2 x^{-\left(1+\frac{2}{\alpha_{s^{\prime}}}\right)} \Lambda_{1,s^{\prime}}^{\prime}([0,x)) \mathrm{d}x. \label{Psi1}
\end{align}

\textit{Proof}: See Appendix \ref{Proof of Lemma 4}. \hfill $\square$

\subsection{Energy Coverage Probability}
The energy harvested at a typical UE in unit time is expressed as $E_k=\xi P_{r,k}$ where $\xi \in (0,1]$ is the
rectifier efficiency, and $P_{r,k}$ is the total received power given in (\ref{total_received_power}).
Since the effect of additive noise power is negligibly small relative to the total received power, it is omitted \cite{Khan}. The conditional energy coverage probability $\EC_k(\Gamma_k)$ is the probability that the harvested energy $E_k$ is larger than the energy outage threshold $\Gamma_k>0$ given that the typical UE is associated with a UAV from the $k^{\thh}$ tier, i.e., $\EC_k(\Gamma_k)= \mathbb{P}(E_k>\Gamma_k|t=k)$. Therefore, total energy coverage probability $\EC$ for the typical UE can be formulated as
\begin{equation}
\EC=\sum_{k=0}^1 \sum_{s \in \{\LOS,\NLOS\}}  \left[\EC_{k,s} (\Gamma_k)\mathcal{A}_{k,s}\right], \label{CoverageProbability}
\end{equation}
where $\EC_{k,s}(\Gamma_k)$ is the conditional energy coverage probability given that the UE is associated with a $k^{\thh}$ tier LOS/NLOS UAV, and $\mathcal{A}_{k,s}$ denotes the association probability. The following theorem provides our main characterization regarding the total energy coverage probability.

\begin{Theorem1} \label{theo:energycoverage} In a UAV network with practical HH antenna radiation patterns and clustered UEs, the total energy coverage probability for the typical UE is approximately given at the top of the next page in (\ref{total_energy_coverage})
\begin{figure*}
\begin{align}
\EC &\approx  \sum_{s \in \{\LOS,\NLOS\}} \sum_{n=0}^{\mathcal{N}}(-1)^{n} {\mathcal{N} \choose n} \nonumber \\
& \times \Bigg[ \int_{H^{\alpha_{s}}}^{\infty}\left(1+\hat{a} P_0H^2 l_{0,s}^{-\left(1+\frac{2}{\alpha_s}\right)}\right)^{-1} \mathcal{L}_{I_{1,0}}\left(\Gamma_0,\mathcal{E}_{0,0}\right) \prod_{m \in \{\LOS,\NLOS\}} \hspace{-0.3cm} \bar{F}_{L_{1,m}}\left(\left(\frac{P_1}{P_0}l_{0,s}^{\frac{2}{\alpha_s}+1}\right)^{\frac{\alpha_m}{\alpha_m+2}}\right) f_{L_{0,s}}(l_{0,s}) \mathrm{d}l_{0,s}  \nonumber \\
&+\int_{H^{\alpha_{s}}}^{\infty} \left(1+\hat{a} P_1H^2 l_{1,s}^{-\left(1+\frac{2}{\alpha_s}\right)}\right)^{-1}  \left(\prod_{j=0}^{1} \mathcal{L}_{I_{j,1}}\left(\Gamma_1,\mathcal{E}_{1,0}\right)\right) \sum_{m \in \{\LOS,\NLOS\}} \hspace{-0.3cm} \bar{F}_{L_{0,m}}\left(\left(\frac{P_0}{P_1}l_{1,s}^{\frac{2}{\alpha_s}+1}\right)^{\frac{\alpha_m}{\alpha_m+2}}\right) \bar{F}_{L_{1,s^{\prime}}}(l_{1,s}) f_{L_{1,s}}(l_{1,s})\mathrm{d}l_{1,s} \Bigg] \label{total_energy_coverage}
\end{align}
\end{figure*}
where $\hat{a}=\frac{n\eta }{\Gamma_k /\xi}$, $\eta=\mathcal{N}(\mathcal{N}!)^{-\frac{1}{\mathcal{N}}}$, $\mathcal{N}$ is the number of
terms in the approximation and the Laplace transforms of the interference terms are given by
\begin{align}
&\mathcal{L}_{I_{0,k}}(\Gamma_k,\mathcal{E}_{k,0}) \nonumber \\
& =\sum_{s^{\prime} \in \{\LOS,\NLOS\}}  \int_{\mathcal{E}_{k,0}}^{\infty} \left(1+\hat{a} P_0 H^2 x^{-\left(1+\frac{2}{\alpha_{s^{\prime}}}\right)} \right)^{-1} f_{L_{0,s^{\prime}}}(x) \mathrm{d}x, \label{LT_I0}
\end{align}
\begin{align}
&\mathcal{L}_{I_{1,k}}(\Gamma_k,\mathcal{E}_{k,0}) \nonumber \\
&  = \prod_{s^{\prime} \in \{\LOS,\NLOS\}} \hspace{-0.3cm} \exp \Bigg (-\int_{\mathcal{E}_{k,0}}^{\infty} \left( 1-\left(1+\hat{a} P_1 H^2 x^{-\left(1+\frac{2}{\alpha_{s^{\prime}}}\right)} \right)^{-1} \right) \nonumber \\
& \times \Lambda_{1,s^{\prime}}^{\prime}([0,x)) \mathrm{d}x\Bigg). \label{LT_I1}
\end{align}
\end{Theorem1}
\emph{Proof:} See Appendix \ref{Proof of Theorem 1}. \hfill $\square$

Note that since $0^{\thh}$ tier consists of only one UAV, i.e., the cluster center UAV, Laplace transform expression
$\mathcal{L}_{I_{0,0}}(\Gamma_0,\mathcal{E}_{0,0})=1$. The total energy coverage probability
of the network in Theorem \ref{theo:energycoverage} is obtained by first computing the conditional energy coverage probability given that a UE is associated with a $k^{\thh}$ tier LOS/NLOS UAV, and then summing up the conditional probabilities weighted with their corresponding association probabilities. In order to formulate the conditional energy coverage probabilities, Laplace transforms of the interference terms are determined. We also note that although the energy coverage probability approximation in Theorem \ref{theo:energycoverage} involves multiple integrals, we explicitly see the dependence of the energy coverage on, for instance, UAV heights,  path loss distributions, path loss exponents, transmission power levels. Moreover, the integrals can be readily computed via numerical integration methods, providing us with additional insight on the impact of key system/network parameters, as demonstrated in the next section.

\section{Extension to a Model with UAVs at Different Heights}
In the preceding analysis, we analyzed the energy coverage performance of a network in which UAVs are located at a height of $H$ above the ground, and $H$ is assumed to be the same for all UAVs. However, the proposed analytical framework can also be employed to analyze the coverage probability when UAV height is not fixed, i.e., UAVs are assumed to be located at different heights. In the extended model, we consider a more general network in which UAVs are located at different heights. Therefore, we assume that there are $M$ groups of UAVs such that the $\mu^{\thh}$ UAV group is located at the height level $H_{\mu}$ for $\mu=1,2,\ldots,M$ and UAVs at each height level can be considered as a UAV-tier distributed according to an independent homogeneous PPP with density of $\lambda_{U,\mu}$ and the total density is equal to $\sum_{\mu=1}^{M} \lambda_{U,\mu}=\lambda_U$. Different from the preceding analysis in which we have considered a single typical UE located at the origin and named its cluster center UAV as $0^{\thh}$ tier UAV, a separate typical UE for each UAV tier needs to be considered in the coverage probability analysis for this model with UAVs at different heights. For example, when we are analyzing the energy coverage probability of the network for a UE clustered around a $\mu^{\thh}$ tier UAV, we assume that the typical UE is located at the origin and its cluster center UAV is considered as the $0^{\thh}$ tier UAV similar to the previous model. Therefore, total energy coverage probability of the network given that the typical UE is clustered around a $\mu^{\thh}$ tier UAV for $\mu=1,2,\ldots,M$ can be computed as follows:
\begin{equation}
\EC_{\mu} \!\!=\!\!\sum_{k=0}^M \sum_{\substack{s \in \{\LOS, \NLOS\}}} \left[\EC_{\mu,k,s} (\Gamma_k)\mathcal{A}_{\mu,k,s}\right], \label{EnergyCoverageProbability_multiheight}
\end{equation}
where $\EC_{\mu,k,s}(\Gamma_k)$ is the conditional energy coverage probability given that the typical UE is clustered around a $\mu^{\thh}$ tier UAV and it is associated with a $k^{\thh}$ tier LOS/NLOS UAV, and $\mathcal{A}_{\mu,k,s}$ is the association probability with a $k^{\thh}$ tier LOS/NLOS UAV.

\begin{Theorem1}
In a UAV network with practical HH antenna radiation patterns and clustered UEs, the total energy coverage probability of the network given that the typical UE is clustered around a $\mu^{\thh}$ tier UAV  is approximately given at the top of the next page in (\ref{total_energy_coverage_multiheight})
\begin{figure*}
\begin{align}
\EC_{\mu} & \approx \sum_{s \in \{\LOS,\NLOS\}} \sum_{n=0}^{\mathcal{N}}(-1)^{n} {\mathcal{N} \choose n} \nonumber \\
& \times \Bigg[ \int_{ H_{\mu}^{\alpha_s}}^{\infty}\left(1+\hat{a} P_0H_{\mu}^2 l_{0,s}^{-\left(1+\frac{2}{\alpha_s}\right)}\right)^{-1} \left(\prod_{j=1}^{M} \mathcal{L}_{I_{j,0}}\left(\Gamma_0,\mathcal{E}_{0,0}\right)\right) \prod_{m \in \{\LOS,\NLOS\}} \prod_{j=1}^{M} \bar{F}_{L_{j,m}}\left(\left(\frac{P_j}{P_0}l_{0,s}^{\frac{2}{\alpha_s}+1}\right)^{\frac{\alpha_m}{\alpha_m+2}}\right) f_{L_{0,s}}(l_{0,s}) \mathrm{d}l_{0,s} \nonumber \\
&+ \sum_{k=1}^{M} \int_{H_{\mu}^{\alpha_{s}}}^{\infty} \left(1+\hat{a} P_kH_k^2 l_{k,s}^{-\left(1+\frac{2}{\alpha_s}\right)}\right)^{-1}  \left(\prod_{j=0}^{M} \mathcal{L}_{I_{j,k}}\left(\Gamma_k,\mathcal{E}_{k,0}\right)\right) \sum_{m \in \{\LOS,\NLOS\}} \hspace{-0.3cm} \bar{F}_{L_{0,m}}\left(\left(\frac{P_0}{P_k}l_{k,s}^{\frac{2}{\alpha_s}+1}\right)^{\frac{\alpha_m}{\alpha_m+2}}\right) \nonumber \\
& \times \prod_{m \in \{\LOS,\NLOS\}} \prod_{j=1,j \neq \mu}^{M} \bar{F}_{L_{j,m}}\left(\left(\frac{P_j}{P_k}l_{k,s}^{\frac{2}{\alpha_s}+1}\right)^{\frac{\alpha_m}{\alpha_m+2}}\right) \bar{F}_{L_{k,s^{\prime}}}(l_{k,s}) f_{L_{k,s}}(l_{k,s})\mathrm{d}l_{k,s} \Bigg] \label{total_energy_coverage_multiheight}
\end{align}
\end{figure*} \normalsize
\end{Theorem1}
\emph{Proof:} Derivation of $\EC_{\mu}$ follows similar steps as that of $\EC$ in (\ref{total_energy_coverage}). In particular, Laplace transforms $\mathcal{L}_{I_{0,k}}$ and $\mathcal{L}_{I_{j,k}}$ for $j=1,2,\ldots,M$ are computed using the Laplace transform equations given in (\ref{LT_I0}) and (\ref{LT_I1}), respectively, by updating UAV height as $H_j$ and UAV density as $\lambda_j$ for $j=0,1,\ldots,M$. $\bar{F}_{L_{0}}(x)$ and $f_{L_{0,s}}(x)$ are computed using (\ref{CCDF_0}) and (\ref{f_L0s}), respectively, by denoting the UAV height as $H_{\mu}$. Furthermore, $\bar{F}_{L_{k,s}}(x)$ and $f_{L_{k,s}}(x)$ are computed using (\ref{CCDF_1}) and (\ref{f_L1s}), respectively, by updating UAV height as $H_k$ and UAV density as $\lambda_k$ for $k=1,\ldots,M$.

%
\section{Simulation and Numerical Results} \label{sec:Simulation and Numerical Results}
In this section, we present the numerical evaluations of theoretical expressions in addition to the simulation results
which are provided to confirm the accuracy of the proposed UAV network model and the analytical characterizations.
In the numerical computations and simulations, unless stated otherwise, the parameter values listed in Table \ref{Table} are used.

\begin{table}
\small
\caption{System Parameters}
\centering
  \begin{tabular}{| p{3.6cm} | p{1.5cm}| p{1.5cm}|}
    \hline
    \footnotesize \textbf{Description}  & \footnotesize \textbf{Parameter}  & \footnotesize \textbf{Value}  \\ \hline
    \scriptsize Path-loss exponents &\scriptsize $\alpha_{\LOS}$, $\alpha_{\NLOS}$ & \scriptsize 2, 4 \\ \hline
    \scriptsize Environment dependent constants & \scriptsize  $b$, $c$ & \scriptsize $11.95$, $0.136$ \\ \hline
    \scriptsize Height of UAVs & \scriptsize $H$  & \scriptsize $50$ m \\ \hline
    \scriptsize Transmit power & \scriptsize $P_k$ $\forall k$ & \scriptsize 37 dBm \\ \hline
    \scriptsize Energy outage threshold &\scriptsize $\Gamma_k$ \scriptsize $\forall k$ &  \scriptsize 0 dBm \\ \hline
    \scriptsize UAV density & \scriptsize $\lambda_U$ & \scriptsize $10^{-4}$ $(1/\text{m}^2)$ \\ \hline
    \scriptsize UE distribution's standard deviation & \scriptsize $\sigma_c$  & \scriptsize 10 \\ \hline
    \scriptsize Rectifier efficiency &  \scriptsize $\xi$  & \scriptsize 1 \\ \hline
  \end{tabular} \label{Table}
\end{table}

\subsection{Impact of Cluster Size}
\begin{figure*}
\centering
\begin{subfigure}{\figsize\textwidth}
  \includegraphics[width=1\textwidth]{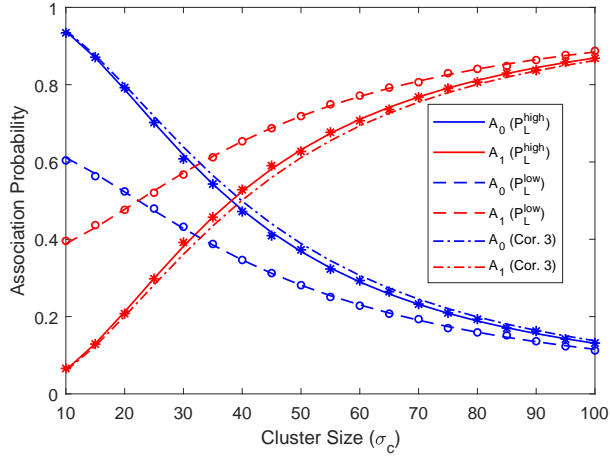}
  \caption{Association probability.}
\label{Fig_AP1}
\end{subfigure}
\newline
\begin{subfigure}[b]{0.48\textwidth}
  \includegraphics[width=1\textwidth]{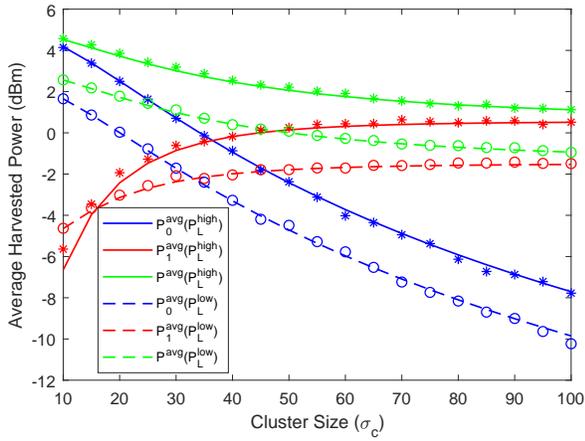}
\caption{Average harvested power.}
\label{Fig_AHP1}
\end{subfigure}
\begin{subfigure}[b]{0.48\textwidth}
  \includegraphics[width=1\textwidth]{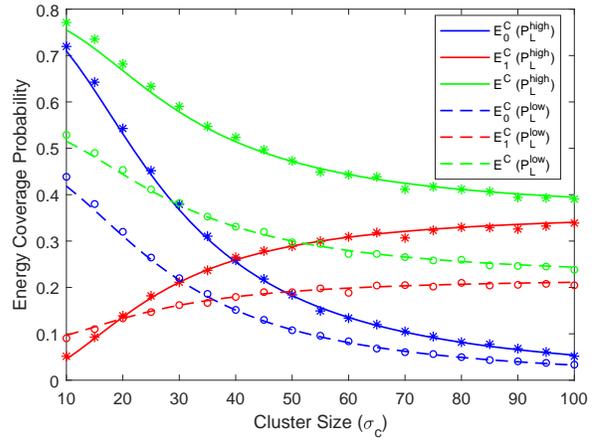}
\caption{Energy coverage probability.}
\label{Fig_ECP1}
\end{subfigure}
\caption{\small (a) Association probability, (b) average harvested power and (c) energy coverage probability as a function of UE distribution's standard deviation $\sigma_c$ for LOS probability functions of high-altitude and low-altitude models when $H=50$ m. Simulation results are plotted with markers while dashed/solid curves show theoretical results. \normalsize}
\end{figure*}

First, we address the effect of UE distribution's standard deviation $\sigma_c$ on the association probability, average harvested power and the energy
coverage probability using the LOS probability functions of high-altitude and low-altitude models of (\ref{LOS_probability1}) and
(\ref{LOS_probability2}) in Figs. \ref{Fig_AP1}, \ref{Fig_AHP1} and \ref{Fig_ECP1}. As the standard deviation increases, the UEs are spread more widely, resulting in generally larger distances between the cluster-center $0^{\thh}$ tier UAV and UEs. For example, for $\sigma_c =10$, the average link distance between cluster center UAV and UEs is around 12.5 meters, while it is around 115 meters for $\sigma_c =90$. Consequently, association probability with the $0^{\thh}$ tier UAV, $\mathcal{A}_{0}$, diminishes, while the association
probability with $1^{\st}$ tier UAVs, $\mathcal{A}_{1}$, increases for both models. Also, for a fixed height, LOS probability of cluster
center UAV decreases for both models with the increasing cluster
size, and hence association probabilities exhibit similar trends. Therefore, the average harvested power from the $0^{\thh}$ tier UAV, $P^{\avg}_{0}$,
increases while the average harvested power from the $1^{\st}$ tier UAVs, $P^{\avg}_{1}$, decreases as the cluster size
grows in both models. On the other hand, the increase in $P^{\avg}_{1}$ cannot compensate the decrease in $P^{\avg}_{0}$, and therefore the total average harvested power $P^{\avg}$ diminishes. In other words, smaller cluster size, i.e., more compactly distributed UEs results in a higher $P^{\avg}$. The energy coverage probability in Fig. \ref{Fig_ECP1} exhibits a very similar behavior as the average harvested power in Fig. \ref{Fig_AHP1}. Also note that, the association probability results of the Corollary 3 closely approximate the association probabilities of the high-altitude model. Finally, we note that simulation results are also plotted in the figure with markers and there is generally an excellent agreement with the analytical results, further validating our analysis.

%

\subsection{Impact of UAV Height} \label{sec:Impact_UAV_height}
\begin{figure}
\centering
\begin{subfigure}{\figsize\textwidth}
\centering
  \includegraphics[width=1\textwidth]{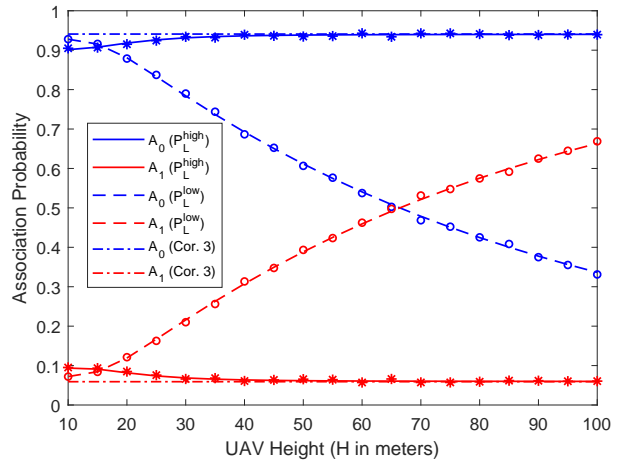}
  \caption{Association probability.}
  \label{Fig_AP2}
\end{subfigure}
\newline
\begin{subfigure}{\figsize\textwidth}
\centering
  \includegraphics[width=1\textwidth]{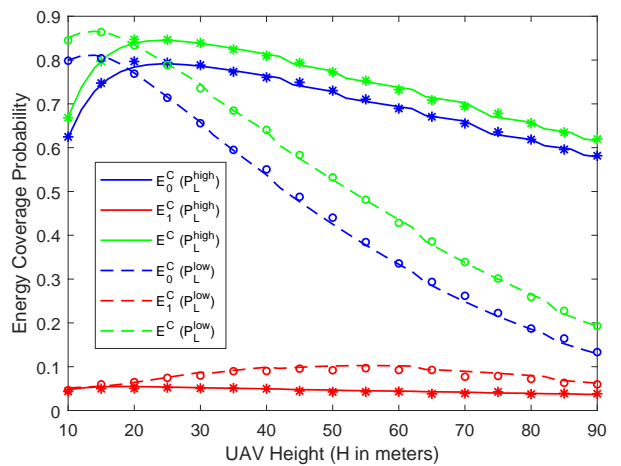}
\caption{Energy coverage probability.}
\label{Fig_ECP2}
\end{subfigure}
\caption{\small (a) Association probability and (b) energy coverage probability as a function of UAV height $H$ for LOS probability functions of
  high-altitude and low-altitude models when $\sigma_c=10$. Simulation results are plotted with markers while dashed/solid curves show theoretical results.  \normalsize}
\end{figure}

Next, in Figs. \ref{Fig_AP2} and \ref{Fig_ECP2}, we plot the association probability and energy coverage probability as a function of UAV height considering the LOS probability functions of both
high-altitude and low-altitude models. For the high-altitude model, since LOS probability increases with the increasing UAV height,
association probability with the $0^{\thh}$ tier UAV increases slightly. On the other hand, LOS probability decreases as a result
of the increase in the 3D distance with the increasing UAV height in the low-altitude model. Therefore, more UEs prefer to connect
to $1^{\st}$ tier UAVs (i.e., UAVs other than the cluster-center one) at higher values of the UAV height. Also note that the result of the Corollary 3  very closely approximates the association probabilities of the high-altitude model especially as the UAV height increases.

Energy coverage probability of the cluster center UAV, $\EC_0$, exhibits similar trends for both types of LOS functions. More specifically, $\EC_0$ increases first then it starts decreasing with the increasing UAV height. Since the effective antenna gain for HH antenna orientation is an increasing function of UAV height for a fixed cluster size, an initial increase in $\EC_0$ is expected. However, further increase in UAV
height results in a decrease in $\EC_0$ of both high-altitude and low-altitude models due to the increase in the distance. Therefore, for
a fixed cluster size, there exists an optimal UAV height maximizing the network energy coverage, $\EC$, for both models. On the other
hand, optimal height maximizing the $\EC$ in the low-altitude model is lower and $\EC$ decreases faster than that in the high-altitude model
because the LOS probability function of the low-altitude model is a decreasing function of distance while the LOS probability function of the high-altitude
model is an increasing function of the UAV height (e.g., as seen in Fig. \ref{Fig_LOS_func}). Moreover, since UEs are more compactly distributed around the
cluster center UAVs for $\sigma_c=10$, energy coverage probability of the $1^{\st}$ tier UAVs, $\EC_1$, is relatively small and changes only very slightly for both models.

%

\subsection{Impact of Antenna Orientation and UAV Density}

\begin{figure*}
    \centering
    \begin{subfigure}[b]{0.47\textwidth}
        \includegraphics[width=\textwidth]{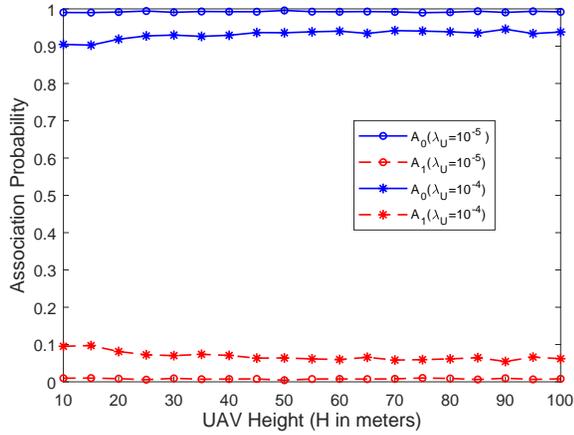}
        \caption{HH orientation.}
        \label{Fig_AP4_1}
    \end{subfigure}
    ~ 
    \begin{subfigure}[b]{0.47\textwidth}
        \includegraphics[width=\textwidth]{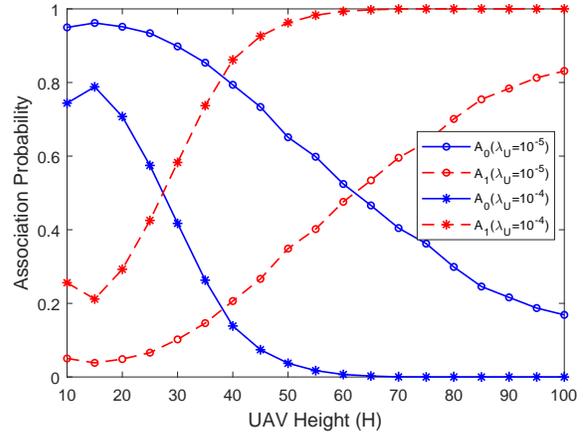}
        \caption{VV orientation.}
        \label{Fig_AP4_2}
    \end{subfigure}
    ~ 
    \\
    \begin{subfigure}[b]{0.5\textwidth}
        \includegraphics[width=\textwidth]{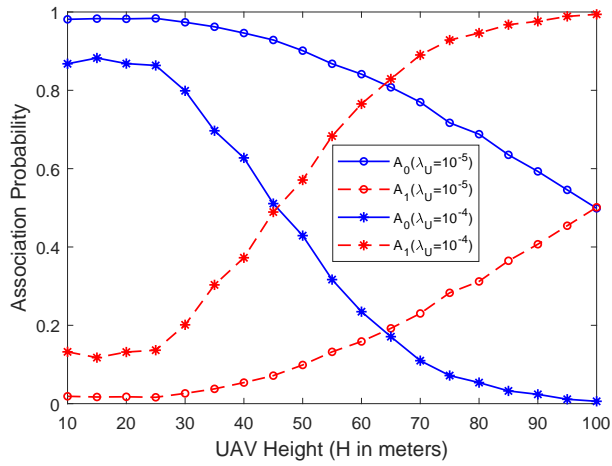}
        \caption{HV orientation.}
        \label{Fig_AP4_3}
    \end{subfigure}
    \caption{\small Association probability as a function of UAV height $H$ for different values of UAV density $\lambda_U$ for
    (a) HH, (b) VV and (c) HV antenna orientations when $\sigma_c=10$. \normalsize}\label{Fig_AP4}
\end{figure*}

\begin{figure}
\centering
  \includegraphics[width=\figsize\textwidth]{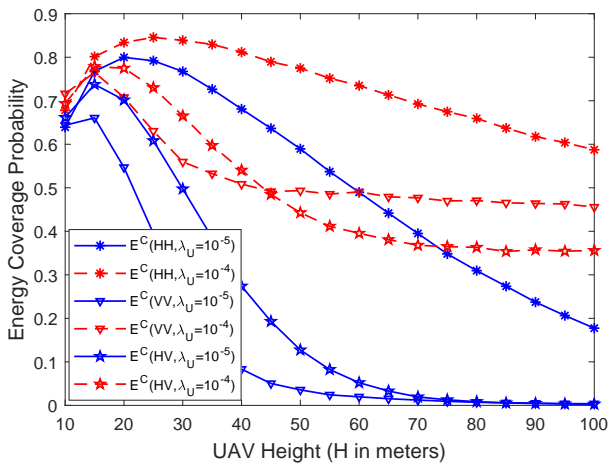}
  \caption{\small Energy coverage probability as a function of UAV height $H$ for different values of UAV density $\lambda_U$  for
  different antenna orientations when $\sigma_c=10$. \normalsize}
\label{Fig_ECP5}
\end{figure}

In Figs. \ref{Fig_AP4_1}, \ref{Fig_AP4_2} and \ref{Fig_AP4_3}, we plot the association probability as a function of UAV height $H$ for different values of UAV density $\lambda_U$ for three different antenna orientations considering the high-altitude LOS probability model. Note that since the analysis for VV and HV antenna orientations seems to be intractable, only simulation
results are plotted. Since effective antenna gain depends on the sine function of the angle between the UAVs and UEs for HH antenna orientation, UEs prefer to connect to their cluster center UAV, and hence $\mathcal{A}_0$ is much larger than $\mathcal{A}_1$ even when there is an increase in the number of UAVs (as seen when the UAV density is increased from $\lambda_U = 10^{-5}$  to $\lambda_U = 10^{-4}$) as shown in Fig. \ref{Fig_AP4_1}. Also note that since both antenna gain and LOS probability is an increasing function with UAV height, increase in
them can compensate the increasing path loss and the association probabilities remain almost constant.


For the VV antenna orientation, effective antenna gain depends on the cosine of the angle between the UAVs and UEs. For larger values of UAV density, association probability with the $0^{\thh}$ tier UAV, $\mathcal{A}_{0}$, slightly increases with increasing UAV height at first as a result of the increase in both the LOS probability and the effective antenna gain. Subsequently, $\mathcal{A}_{0}$ starts decreasing because the increase in the LOS probability cannot compensate the rapid decrease in the effective antenna gain between the UE and the cluster center UAV. For a less dense network, UEs associate with the cluster-center UAV mostly at lower UAV heights. However, with the increasing height, antenna gain with the cluster-center UAV decreases and consequently, the association probability with $1^{\st}$ tier UAVs, $\mathcal{A}_1$, increases.

Finally, for the HV antenna orientation, effective antenna gain is a function of both cosine and sine of the angle $\theta$. Association probabilities exhibit similar trends as in the case of VV orientation. However, different from the VV case, since the antenna gain depends on both the cosine and sine functions, association probability with the $0^{\thh}$ tier UAV, $\mathcal{A}_{0}$, is greater than that of the VV orientation.

We also plot the energy coverage probability for different UAV heights, antenna orientations, and UAV densities in Fig. \ref{Fig_ECP5}. The performance with the HV and VV antenna orientations exhibit similar behaviors as that with the HH antenna orientation which is described in Section \ref{sec:Impact_UAV_height}. For both higher-density (i.e., $\lambda_U = 10^{-4}$) and lower-density UAV networks (i.e., $\lambda_U = 10^{-5}$), HH orientation leads the best performance compared to the HV and VV cases. We also note that as the UAV density and UAV height are increased, energy coverage probability with the VV orientation starts exceeding that of the HV case. Therefore, energy coverage performance can be improved by varying the antenna orientations depending on the number of UAVs in the network and their height.

\begin{figure}
\centering
  \includegraphics[width=\figsize\textwidth]{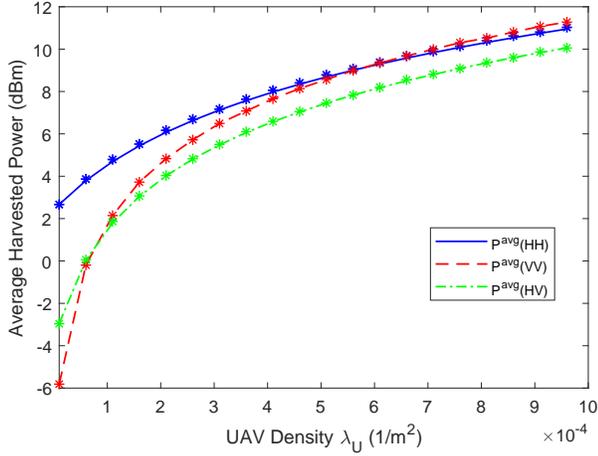}
  \caption{\small Average harvested power as a function of UAV density $\lambda_U$. \normalsize}
\label{Fig_AHP_lambdaU}
\end{figure}

\begin{figure}
\centering
  \includegraphics[width=\figsize\textwidth]{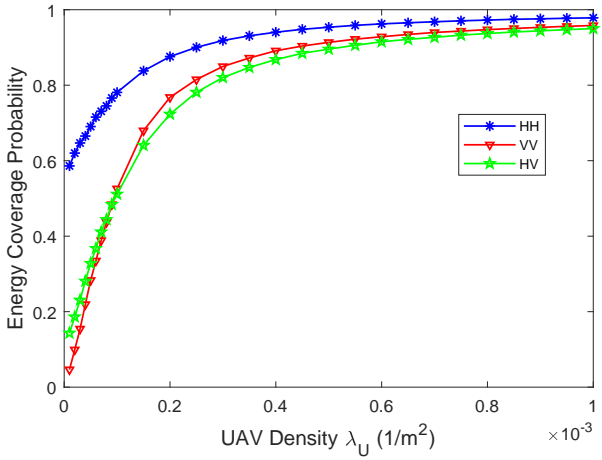}
  \caption{\small Energy coverage probability as a function of UAV density $\lambda_U$. \normalsize}
\label{Fig_IG_afo_lambdaU2}
\end{figure}

Furthermore, we display the average harvested power and the energy coverage probability as a function of the UAV density for three different antenna orientations considering the high-altitude LOS probability model in Figs. \ref{Fig_AHP_lambdaU} and \ref{Fig_IG_afo_lambdaU2}, respectively. Both the average harvested power and the energy coverage probability are increasing functions of the UAV density irrespective of the antenna orientation for a fixed UAV height. Expectedly, adding more UAVs to the network results in an increase in the total power received at the typical UE, and hence the average harvested power grows and the energy coverage performance of the network improves. We also note that HH antenna orientation generally leads to larger average harvested power and energy coverage probability. On the other hand, VV antenna orientation results in a higher average harvested power when the UAV density is sufficiently large due to the fact that one can harvest more power from the dense $1^{\text{st}}$-tier UAVs with smaller elevation angles when this antenna orientation is used.

\subsection{Impact of UAV Transmit Power}
\begin{figure}
\centering
  \includegraphics[width=\figsize\textwidth]{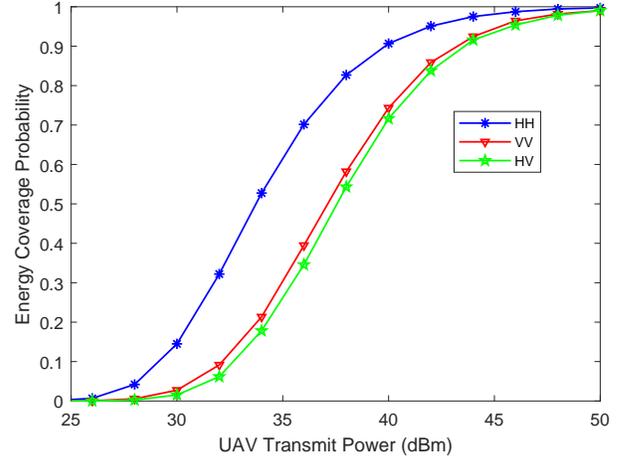}
  \caption{\small Energy coverage probability as a function of UAV transmit power. \normalsize}
\label{Fig_ECP_afo_tpower}
\end{figure}

In Fig. \ref{Fig_ECP_afo_tpower}, we plot the energy coverage probability as a function of the UAV transmit power for three different antenna orientations considering the high-altitude LOS probability model. As expected, energy coverage probability is an increasing function of UAV transmit power. Moreover, HH antenna orientation generally results in a higher energy coverage probability than VV and HV antenna orientations.

\subsection{Impact of Energy Outage Threshold}
\begin{figure}
\centering
  \includegraphics[width=\figsize\textwidth]{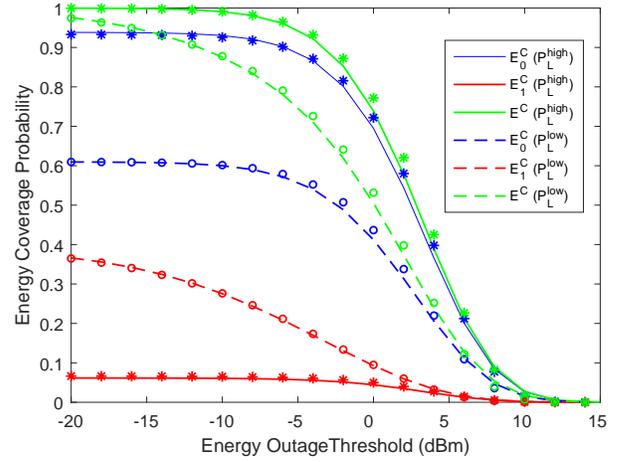}
  \caption{\small Energy coverage probability as a function of energy outage threshold in dBm for LOS probability functions of high-altitude and low-altitude models when $\sigma_c=10$ and $H=50$ m. Simulation results are plotted with markers while dashed/solid curves show theoretical results.  \normalsize}
\label{Fig_ECP}
\end{figure}

In Fig. \ref{Fig_ECP}, we show the energy coverage probabilities of different tiers (i.e., $\EC_0$ and $\EC_1$) and the total energy coverage probability $\EC$ as a function of the energy outage threshold for both high-altitude and low-altitude models. As seen in Fig. \ref{Fig_AP1} and Fig. \ref{Fig_AP2}, UEs are more likely to be associated with the $0^{\thh}$ tier UAV rather than $1^{\st}$ tier UAVs in the high-altitude model when $\sigma_c=10$, and hence $\EC_0$ is much higher than $\EC_1$. On the other hand, for the low-altitude model, since association probabilities with each tier are not very different, more UEs can be covered by $1^{\st}$ tier UAVs compared to the high-altitude model. However, $\EC_0$ is still greater than $\EC_1$ due to the relatively smaller distance to the cluster-center UAV. We also observe that as a general trend, energy coverage probabilities expectedly diminish with increasing energy outage threshold.

\subsection{Impact of Different UAV Heights}
\begin{figure}
\centering
  \includegraphics[width=\figsize\textwidth]{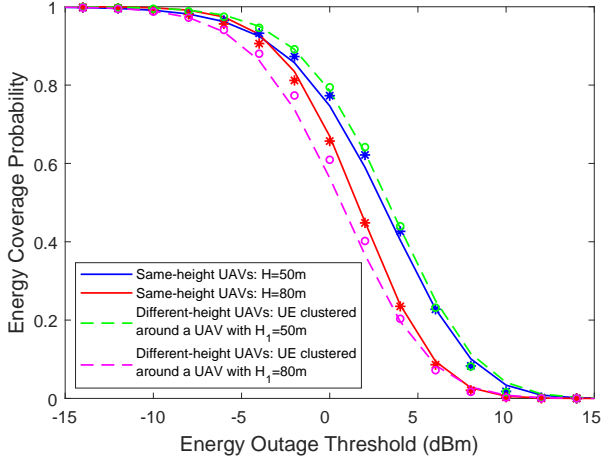}
  \caption{\small Energy coverage probability as a function of energy outage threshold in dBm for high-altitude LOS probability function when $\sigma_c=10$. Solid lines show the energy coverage probabilities when half of the UAVs are located at height $H_1 = 50$ m and the other half are located at height  $H_2 = 80$ m, and the typical UE is clustered around a UAV at either height $H_1$ or $H_2$. Simulation results are also plotted with markers.  \normalsize}
\label{Fig_ECP_multiheight}
\end{figure}

Finally, in Fig. \ref{Fig_ECP_multiheight}, we plot the total energy coverage probabilities as a function of the energy outage threshold using the high-altitude LOS probability function model when UAVs are assumed to be located at different heights. In this setup, we use the same parameters given in Table \ref{Table} with some differences in UAV height and UAV density. More specifically, we consider $M = 2$ groups of UAVs located at altitudes $H_1 = 50$ m and $H_2 = 80$ m with densities $\lambda_{U,1} = \lambda_{U,2}=\lambda_{U}/2$ and transmit powers $P_1 = P_2 = 37$ dBm. Therefore, transmit power of the $0^{\thh}$ UAV is also equal to $P_0 = 37$ dBm. Similar to \cite{Liu}, we consider practical values for UAV heights which are around $50\sim100$ meters.

In Fig. \ref{Fig_ECP_multiheight}, solid lines plot the total energy coverage probabilities when the height is the same for all UAVs. Dashed lines display the coverage probabilities when half of the UAVs are located at height $H_1$ and the other half are located at height $H_2$, and the typical UE is clustered around a UAV at either height $H_1$ or $H_2$. As seen in the figure, energy coverage performance of the different-height UAV network is very similar to that of the same-height UAV. For example, if a typical UE is clustered around a UAV at height $H_1 = 50$ m, the performance of the different-height UAV network is almost the same with that of the same-height UAV for $H=50$ m. On the other hand, performance of the different-height network when the typical UE is clustered around a UAV at height $H_1 = 80$ m is slightly lower than that of the same-height UAV network with $H=80$ m as a result of growing impact of interferences from the UAVs at height $H_2=50$ m.

\section{Conclusion} \label{sec:Conclusion}
In this paper, we have analyzed the energy coverage probability of a UAV network with clustered UEs.
The UEs are distributed according to a PCP around PPP distributed cluster centers. The UAVs are deployed at a certain height above these cluster centers, which, as a result, represent the projections of
UAVs on the ground. UEs are assumed to be associated with the UAV providing the strongest long-term averaged power. In this setting, we have determined the association probabilities and characterized the energy coverage probability. We have analyzed the effect of two different LOS probability functions on the network performance. We have also investigated the impact of practical 3D antenna radiation patterns on the network performance. Furthermore, we have presented energy coverage probability expressions for a more general model in which UAVs are located at different heights.

Via numerical results, we have demonstrated that the standard deviation of UE distribution $\sigma_c$, UAV height $H$, and antenna orientation
have considerable influence on UAV association and energy coverage probabilities. For instance, more widely spread UEs result in a  decrease in the total energy coverage probability of the network for both LOS probability models. For a certain cluster size, there exists an optimal UAV height that maximizes the network energy coverage. However, this optimal height depends on the type of the LOS probability model. In particular, low-altitude model features a lower optimal height than the high-altitude model. Although the maximum value of the energy coverage reached at the optimal height is almost the same for both models, low-altitude model has better energy coverage performance at lower heights. On the other hand, high-altitude model performs better at higher heights. These observations indicate that energy coverage performance is greatly affected by the environment. We have also shown that antenna orientation has a significant impact on the energy coverage probability depending on the UAV density. Specifically, better performance can be achieved by changing the antenna orientations according to the number of UAVs in the network and their height. We have also evaluated the average harvested power levels for different antenna orientations at various UAV density levels and computed the energy coverage probability as UAV density, UAV transmission power or energy outage threshold varies and we identified the impact of these different settings on the performance. Finally, we have addressed the performance of the different-height network and noted that UAVs at lower heights can lead to increased interference. Analyzing the performance of a UAV network with simultaneous information and energy transfer remains as future work.

\appendix
\subsection{Proof of Lemma 1}
\label{Proof of Lemma 1}
The CCDF of the path loss $L_{0,s}$ from a typical UE to a $0^{th}$ tier LOS/NLOS UAV is
\begin{align}
&\bar{F}_{L_{0,s}}(x) \nonumber \\
&= \mathbb{E}_r \left[\mathbb{P}\left(L_{0,s}(r)\geq x \right) \mathcal{P}_s(r)\right] \nonumber \\
&=  \mathbb{E}_d \left[\mathbb{P}\left( (d^2+H^2)^{\alpha_{s}/2}\geq x \right) \mathcal{P}_s(\sqrt{d^2+H^2}) \right]\label{Aeq1} \\
&= \int_{0}^{\infty} \mathbb{P}\left(d \geq \sqrt{x^{2/\alpha_{s}}-H^2} \right) \mathcal{P}_s(\sqrt{d^2+H^2}) f_D(d) \mathrm{d}d \nonumber \\
&= \int_{\sqrt{x^{2/\alpha_{s}}-H^2}}^{\infty} \mathcal{P}_s(\sqrt{d^2+H^2}) f_D(d) \mathrm{d}d
\end{align}
$\text{for } s \in \{\LOS, \NLOS\}$ where $f_{D}(d)$ is given in (\ref{PDF_of_d}), and $\mathcal{P}_s(\cdot)$ is the LOS or NLOS probability depending on whether $s = \LOS$ or $s = \NLOS$. With this, the CCDF of the path loss $L_{0}$ from a typical UE to a $0^{\thh}$ tier UAV given in Lemma 1 can be obtained by summing up over $s$.
\subsection{Proof of Lemma 2}
\label{Proof of Lemma 2}
Intensity function for the path loss model from a typical UE to a $1^{\st}$ tier UAV for $s \in \{\LOS,\NLOS\}$ can be formulated as follows:
\begin{align}
\Lambda_{1,s}([0,x))&=\int_{\mathbb{R}^2} \mathbb{P}\left(L_1(r)<x\right)\mathrm{d}r \label{Aeq3} \\
&=2\pi\lambda_U \int_H^{\infty}\mathbb{P}\left(r^{\alpha_{s}}<x\right) \mathcal{P}_{s}(r) r \mathrm{d}r \label{Aeq31} \\
&= 2\pi\lambda_U \int_H^{\infty} \mathbb{P}\left(r< x^{1/\alpha_{s}}\right)\mathcal{P}_{s}(r) r \mathrm{d}r \\
&=  2\pi\lambda_U \int_{H}^{x^{1/\alpha_{s}}} \mathcal{P}_{s}(r) r \mathrm{d}r \label{app:Lemma1}
\end{align}
where (\ref{Aeq3}) is due to the definition of intensity function for the point process of the path loss. CCDF of the path loss $L_{1}$ from a typical UE to a $1^{\st}$ tier UAV given in Lemma 2 can be obtained by summing up $\Lambda_{1,s}([0,x))$ over $s$.

\subsection{Proof of Lemma 3}
\label{Proof of Lemma 3}
Association probability with a $0^{\thh}$ tier LOS/NLOS UAV can be expressed as
\begin{align}
\mathcal{A}_{0,s} &=\prod_{m \in \{\LOS,\NLOS\}} \mathbb{P}(P_0 G_0(r) L_{0,s}^{-1} \geq P_1 G_1(r) L_{1,m}^{-1}) \label{Aeq7} \\
&= \prod_{m \in \{\LOS,\NLOS\}} \mathbb{P}\left(P_0 \frac{H^2}{L_{0,s}^{\frac{2}{\alpha_s}}} L_{0,s}^{-1} \geq P_1 \frac{H^2}{L_{1,m}^{\frac{2}{\alpha_L}}} L_{1,m}^{-1}\right)  \label{Aeq71} \\
&= \prod_{m \in \{\LOS,\NLOS\}} \mathbb{P}\left( L_{1,m} \geq \left(\frac{P_1}{P_0}L_{0,s}^{\frac{2}{\alpha_s}+1}\right)^{\frac{\alpha_m}{\alpha_m+2}} \right) \nonumber \\
&=\int_{ H^{\alpha_{s}}}^{\infty} \prod_{m \in \{\LOS,\NLOS\}} \bar{F}_{L_m}\left(\left(\frac{P_1}{P_0}l_{0,s}^{\frac{2}{\alpha_s}+1}\right)^{\frac{\alpha_m}{\alpha_m+2}}\right) \nonumber \\
& \times f_{L_{0,s}}(l_{0,s}) \mathrm{d}l_{0,s}\label{app:Lemma3_0}
\end{align}
where (\ref{Aeq7}) uses the fact that LOS and NLOS links in the $1^{\st}$ tier are independent, and (\ref{app:Lemma3_0}) incorporates the definition of the CCDF of the path loss. Since the distance between UEs and UAVs is at least $H$, the lower limit of the integration is $l_{0,s}= H^{\alpha_{s}}$.

Association probability with a $1^{\st}$ tier LOS/NLOS UAV is given by
\begin{align}
&\mathcal{A}_{1,s} \nonumber \\
&= \mathbb{P} (L_{1,s^{\prime}}>L_{1,s}) \hspace{-0.3cm} \prod_{m \in \{\LOS,\NLOS\}} \hspace{-0.3cm}  \mathbb{P}(P_1 G_1(r) L_{1,s}^{-1} \geq P_0 G_0(r) L_{0,m}^{-1}) \label{Aeq4} \\
&= \mathbb{P} (L_{1,s^{\prime}}>L_{1,s}) \hspace{-0.3cm} \prod_{m \in \{\LOS,\NLOS\}} \hspace{-0.3cm} \mathbb{P}\left(P_1 \frac{H^2}{L_{1,s}^{\frac{2}{\alpha_s}}} L_{1,s}^{-1} \geq P_0 \frac{H^2}{L_{0,m}^{\frac{2}{\alpha_m}}} L_{0,m}^{-1}\right) \nonumber \\
&=\mathbb{P} (L_{1,s^{\prime}}>L_{1,s}) \hspace{-0.3cm} \prod_{m \in \{\LOS,\NLOS\}} \hspace{-0.3cm} \mathbb{P}\left( L_{0,m} \geq \left(\frac{P_0}{P_1}L_{1,s}^{\frac{2}{\alpha_s}+1}\right)^{\frac{\alpha_m}{\alpha_m+2}} \right)  \nonumber \\
&=\int_{ H^{\alpha_{s}}}^{\infty} \bar{F}_{L_{1,s^{\prime}}}(l_{1,s}) \hspace{-0.3cm} \prod_{m \in \{\LOS,\NLOS\}} \hspace{-0.3cm} \bar{F}_{L_{0,m}}\left(\left(\frac{P_0}{P_1}l_{1,s}^{\frac{2}{\alpha_s}+1}\right)^{\frac{\alpha_m}{\alpha_m+2}}\right) \nonumber \\
& \times f_{L_{1,s}}(l_{1,s})\mathrm{d}l_{1,s}, \label{app:Lemma3_1}
\end{align}
where $s, s' \in \{\LOS,\NLOS\}$, and $s \neq s'$. (\ref{Aeq4}) makes use of the definition of the association probability and the fact that LOS and NLOS links in the $0^{\thh}$ tier are independent, and $\mathbb{P} (L_{1,s^{\prime}}>L_{1,s})=\bar{F}_{L_{1,s^{\prime}}}(l_{1,s})$.

\subsection{Proof of Lemma 4}
\label{Proof of Lemma 4}
The total average harvested power at a typical UE can be formulated as
\begin{align}
P^{\avg}=\sum_{k=0}^{1} \sum_{s \in \{\LOS,\NLOS\}} \left [ P^{\avg}_{k,s} \mathcal{A}_{k,s}\right], \label{app:Lemma4_3}
\end{align}
where $P^{\avg}_{k,s}$ is the conditional average harvested power given that the UE is associated with a LOS/NLOS UAV in tier $k\in\{0,1\}$, and $\mathcal{A}_{k,s}$ denotes the association probability. Conditional average harvested power can be obtained as follows:
\begin{align}
 P^{\avg}_{k,s} & = \mathbb{E}_{S_{k,s},I_{tot}} \left[ S_{k,s} + \sum_{j=0}^{1} I_{j,k}\right] \nonumber \\
 &= \mathbb{E}_{L_{k,s}}\Bigg[ P_k H^2 L_{k,s}^{-\left(1+\frac{2}{\alpha_s}\right)}\nonumber \\
 &+ \sum_{s^{\prime} \in \{\LOS,\NLOS\}} \int_{\mathcal{E}_{k,0}}^{\infty} P_0 H^2 x^{-\left(1+\frac{2}{\alpha_{s^{\prime}}}\right)}f_{L_{0,s^{\prime}}}(x) \mathrm{d}x \nonumber \\
 & + \sum_{s^{\prime} \in \{\LOS,\NLOS\}} \int_{\mathcal{E}_{k,0}}^{\infty} P_1 H^2 x^{-\left(1+\frac{2}{\alpha_{s^{\prime}}}\right)}\Lambda_{1,s^{\prime}}^{\prime}([0,x))\mathrm{d}x \Bigg] \label{app:Lemma4_0} \\
 &=\mathbb{E}_{L_{k,s}}\left[ P_k H^2 L_{k,s}^{-\left(1+\frac{2}{\alpha_s}\right)}+ \sum_{j=0}^{1}\Psi_{I_{j,k}}(\mathcal{E}_{k,0}) \right] \label{app:Lemma4_1}
\end{align}
where (\ref{app:Lemma4_0}) follows from the averaging over the fading distribution, inserting the antenna gain $G_k=H^2 L_{k,s}^{\frac{2}{\alpha_s}}$ for HH antenna orientation and employing Campbell's theorem, (\ref{app:Lemma4_1}) follows from the definitions of $\Psi_{I_{0,k}}$ and $\Psi_{I_{1,k}}$ provided in (\ref{Psi0}) and (\ref{Psi1}), respectively. Note that the interfering $0^{\thh}$ tier UAV and $1^{\st}$ tier UAVs lie outside the exclusion disc $\mathcal{E}_{k,0}$ with radius $\left( \frac{P_0}{P_k}l_{k,s}^{1+\frac{2}{\alpha_s}}\right)^{\frac{\alpha_{s^{\prime}}}{\alpha_{s^{\prime}}+2}}$ and $\mathcal{E}_{k,0}$ with radius $\left( \frac{P_1}{P_k}l_{k,s}^{1+\frac{2}{\alpha_s}}\right)^{\frac{\alpha_{s^{\prime}}}{\alpha_{s^{\prime}}+2}}$, respectively. Finally, by inserting (\ref{Association_Prob0}), (\ref{Association_Prob1}), (\ref{app:Lemma4_1}) into (\ref{app:Lemma4_3}), the average harvested power expression in (\ref{lemma4}) can be obtained.

\subsection{Proof of Theorem 1}
\label{Proof of Theorem 1}
Given that the UE is associated with a LOS/NLOS UAV in tier $k\in\{0,1\}$, the conditional energy coverage probability $\EC_{k,s}(\Gamma_k)$ is
\begin{align}
&\EC_{k,s}(\Gamma_k) \nonumber \\
&=\mathbb{P}(\xi \left( S_{k,s} +I_{tot}\right) >\Gamma_k) \label{ATh11} \\
&\approx \sum_{n=0}^{\mathcal{N}}(-1)^n {\mathcal{N} \choose n} \mathbb{E}_{S_{k,s},I_{tot}}\left[ e^{-\hat{a}(S_{k,s}+I_{tot})}\right] \label{ATh12} \\
&=\sum_{n=0}^{\mathcal{N}}(-1)^n {\mathcal{N} \choose n} \mathbb{E}_{S_{k,s}}\left[ e^{-\hat{a}S_{k,s}} \mathbb{E}_{I_{tot}|S_{k,s}}\left[ e^{-\hat{a}I_{tot}} \right] \right] \nonumber \\
&=\sum_{n=0}^{\mathcal{N}}(-1)^n {\mathcal{N} \choose n} \mathbb{E}_{L_{k,s}} \Bigg[ \left(1+\hat{a}P_kG_kL_{k,s}^{-1}\right)^{-1}  \nonumber \\
& \times \prod_{j=0}^{1}\mathbb{E}_{I_{j,k}|L_{k,s}}\left[ e^{-\hat{a}I_{j,k}}\right]\Bigg] \label{Ath13} \\
&=\sum_{n=0}^{\mathcal{N}}(-1)^n {\mathcal{N} \choose n} \mathbb{E}_{L_{k,s}} \Bigg[ \left(1+\hat{a}P_k H^2 L_{k,s}^{-\left(1+\frac{2}{\alpha_s} \right)}\right)^{-1}  \nonumber \\
& \times \prod_{j=0}^{1}\mathcal{L}_{I_{j,k}}(\Gamma_k,\mathcal{E}_{k,0}) \Bigg] \label{app:Theorem1}
\end{align}
where $\hat{a}=\frac{n\eta }{\Gamma_k /\xi}$, $\eta=\mathcal{N}(\mathcal{N}!)^{-\frac{1}{\mathcal{N}}}$, $\mathcal{N}$ is the number of
terms in the approximation, $\mathcal{L}_{I_{j,k}}(\Gamma_k,\mathcal{E}_{k,0})$ is the Laplace transform of $I_{j,k}$, and (\ref{ATh12}) is approximated by following the similar steps in \cite{Khan}. In (\ref{Ath13}), we inserted the antenna gain $G_k=H^2 L_{k,s}^{\frac{2}{\alpha_s}}$, and the last step in (\ref{app:Theorem1}) follows from $h_{k,0}$ $\sim$ $\exp(1)$ and by noting that Laplace transforms of interference at the UE from different tier UAVs are independent.

Laplace transforms can be determined by employing key characterizations from stochastic geometry. Recall that $0^{\thh}$ tier interference arises from the UAV at the cluster center of the typical UE. When the typical UE is associated with a UAV in the $1^{\st}$ tier, Laplace transform of the interference from $0^{\thh}$ tier UAV can be formulated as
\begin{align}
&\mathcal{L}_{I_{0,k}}(u) \nonumber \\
&\hspace{-0.25cm} = \mathbb{E}_{I_{0,k}}\left[e^{-\hat{a}I_{0,k}}\right]\nonumber\\
& \hspace{-0.25cm}  = \hspace{-0.4cm} \sum_{s^{\prime} \in \{\LOS,\NLOS\}} \hspace{-0.3cm} \mathbb{E}_x\left[\mathbb{E}_{h_{0,0}}\left[\exp\left(-\hat{a} P_0G_0h_{0,0}x^{-1}\right)|P_0G_0x^{-1}<P_kG_kl_k^{-1}  \right] \right] \label{Aeq9} \\
&\hspace{-0.25cm}=\hspace{-0.4cm} \sum_{s^{\prime} \in \{\LOS,\NLOS\}}\hspace{-0.3cm} \mathbb{E}_x\left[\left(1+\hat{a} P_0H^2x^{-\left(1+\frac{2}{\alpha_{s^{\prime}}}\right)}\right)^{-1}\middle |x>\left( \frac{P_0}{P_k}l_{k,s}^{1+\frac{2}{\alpha_s}}\right)^{\frac{\alpha_{s^{\prime}}}{\alpha_{s^{\prime}}+2}}  \right] \label{Aeq10} \\
&\hspace{-0.25cm}=\hspace{-0.4cm}\sum_{s^{\prime} \in \{\LOS,\NLOS\}}  \int_{\mathcal{E}_{k,0}}^{\infty} \left(1+\hat{a} P_0 H^2 x^{-\left(1+\frac{2}{\alpha_{s^{\prime}}}\right)} \right)^{-1}  f_{L_{0,s^{\prime}}}(x) \mathrm{d}x
\end{align} \normalsize
where conditioning in (\ref{Aeq9}) is a result of the UAV association policy, i.e., the received power from the interfering $0^{\thh}$ tier UAV is less than the received power from the associated UAV, (\ref{Aeq10}) follows from $h_{0,0}$ $\sim$ $\exp(1)$ and inserting the antenna gains, in the last step the exclusion disc $\mathcal{E}_{k,0}=\left( \frac{P_0}{P_k}l_{k,s}^{1+\frac{2}{\alpha_s}}\right)^{\frac{\alpha_{s^{\prime}}}{\alpha_{s^{\prime}}+2}}$. Also note that $\mathcal{L}_{I_{0,k}}(u) = 1$, if the typical UE is associated with $0^{\thh}$ tier UAV.

Laplace transform of the interference from $1^{\st}$ tier UAVs is
\begin{align}
&\mathcal{L}_{I_{1,k}}(u)= \mathbb{E}_{I_{1,k}}\left[e^{-\hat{a}I_{1,k}}\right] \nonumber \\
&=\prod_{s^{\prime} \in \{\LOS,\NLOS\}} \exp\Bigg(-\int_{\mathcal{E}_{k,0}}^{\infty}\!\!\!\left(1-\mathbb{E}_{h_{1,i}} \left[ e^{-\hat{a} P_1 H^2 h_{1,i} x^{-\left(1+\frac{2}{\alpha_{s^{\prime}}}\right)} }\right]\right) \nonumber \\
& \times \Lambda_{1,s^{\prime}}^{\prime}([0,x))\mathrm{d}x \Bigg) \label{Aeq11} \\
&=\prod_{s^{\prime} \in \{\LOS,\NLOS\}} \hspace{-0.3cm} \exp \Bigg (-\int_{\mathcal{E}_{k,0}}^{\infty} \left( 1-\left(1+\hat{a} P_1 H^2 x^{-\left(1+\frac{2}{\alpha_{s^{\prime}}}\right)} \right)^{-1} \right) \nonumber \\
& \times \Lambda_{1,s^{\prime}}^{\prime}([0,x)) \mathrm{d}x\Bigg) \label{Aeq12}
\end{align}
where (\ref{Aeq11}) is obtained by computing the probability generating functional of the PPP, and (\ref{Aeq12}) follows by computing the moment generating function of the exponentially distributed random variable $h$. Note that the interfering $1^{\st}$ tier UAVs lie outside the exclusion disc $\mathcal{E}_{k,0}$ with radius $\left( \frac{P_1}{P_k}l_{k,s}^{1+\frac{2}{\alpha_s}}\right)^{\frac{\alpha_{s^{\prime}}}{\alpha_{s^{\prime}}+2}}$. Finally, by inserting (\ref{Association_Prob0}), (\ref{Association_Prob1}), (\ref{LT_I0}), (\ref{LT_I1}) into (\ref{CoverageProbability}), energy coverage probability expression in (\ref{total_energy_coverage}) can be obtained.



\end{document}